\documentclass[twocolumn, aps]{revtex4-2}
\usepackage{graphicx}
\usepackage{amsmath}
\usepackage{graphics}

\usepackage{amsmath}
\usepackage{amsfonts}
\usepackage{amssymb}
\usepackage{xcolor}
\usepackage{subcaption}
\DeclareUnicodeCharacter{2212}{-}

\begin{document}

\title{Emergence of negative viscosities and colored noise under current-driven Ehrenfest molecular dynamics}

\author{Riley J. Preston}
\affiliation{College of Science and Engineering, James Cook University, Townsville, QLD, 4811, Australia }

\author{Thomas D. Honeychurch}
\affiliation{College of Science and Engineering, James Cook University, Townsville, QLD, 4811, Australia }

\author{ Daniel S. Kosov}
\affiliation{College of Science and Engineering, James Cook University, Townsville, QLD, 4811, Australia }

\begin{abstract}
Molecules in molecular junctions are subject to current-induced forces that can break chemical bonds, induce reactions, destabilize molecular geometry, and halt the operation of the junction. Theories behind current-driven molecular dynamics simulations rely on a perturbative time-scale separation within the system with subsequent use of nonequilibrium Green's functions (NEGF) to compute conservative, non-conservative, and stochastic forces exerted by electrons on nuclear degrees of freedom. We analyze the effectiveness of this approximation, paying particular attention to the phenomenon of negative viscosities. The perturbative approximation is directly compared to the nonequilibrium Ehrenfest approach. We introduce a novel time-stepping approach to calculate the forces present in the Ehrenfest method via exact integration of the equations of motion for the nonequilibrium Green's functions, which does not necessitate a time-scale separation within the system and provides an exact description for the corresponding classical dynamics. We observe that negative viscosities are not artifacts of a perturbative treatment but also emerge in Ehrenfest dynamics. However, the effects of negative viscosity have the possibility of being overwhelmed by the predominantly positive dissipation due to the higher-order forces unaccounted for by the perturbative approach. Additionally, we assess the validity of the white-noise approximation for the stochastic forces, finding that it is {  justifiable in the presence of a clear time-scale separation and is more applicable  when the current-carrying molecular orbital is moved outside of the voltage window.} Finally, we demonstrate the method for molecular junction models consisting of one and two classical degrees of freedom.

\end{abstract}

\maketitle

\newpage
\section{INTRODUCTION}

{  The prototypical building block of nanoscale electronics technology is the molecular junction, in which a chosen molecule is bonded between two conducting leads. Current densities in nanoscale electronic devices far outweigh that of their macroscopic counterparts\cite{cuevasbook}. This results in joule heating\cite{heating11,Sabater15,Tsutsui08,Schulze08,Ioffe08,Franke12,Mukherjee13,Huang06,deLeon08,Huttel09,Stettenheim10,Schulze08_2} whereby energy from the tunneling electrons is transferred to the vibrational degrees of freedom of the molecule.} This can result in a plethora of technologically unwanted but physically interesting effects such as large-scale conformational changes of the molecule which affect the conduction properties of the system\cite{Hahn05,ratner13}, as well as total bond rupture and device breakdown\cite{Hla00,Lauhon00,Stipe97,Haixing15,Haixing16,Pobelov17,Sabater15}. Clearly, it is desirable for us to be able to predict and even manipulate such behaviours to our advantage, emphasizing the need for reliable theoretical models which accurately describe the interplay between electronic and vibrational degrees of freedom.

The prevailing theory can be partitioned into two main archetypes; methods which opt to treat the entire system on a quantum footing, and quasi-classical methods which utilise a classical description of nuclei interacting with a quantum environment. Purely quantum models of the molecular dynamics are often limited to harmonic vibrations\cite{Ding2016,Erpenbeck16,thoss14,belzig10,galperin09} or other generic vibrational potentials\cite{Erpenbeck2019,koch06} which are unaffected by the electronic environment. Other quantum methods which are able to accurately account for the large-scale motion required to study device breakdown are generally limited to highly simplistic systems with minimal relevance to experiment or are computationally infeasible to simulate\cite{rabani14,wang11,Schinabeck16}. It is instead convenient to adopt a quasi-classical approach in which the nuclei are assumed to behave according to classical equations of motion under the effects of the quantum tunneling electrons. This enables the description of arbitrary molecular potentials for systems along with being able to capture experimentally observed phenomena such as current-induced heating\cite{preston2020,prestonAC2020,kershaw20,pistolesi16,fuse,Lu15} along with bond rupture and electronically induced chemical reactions\cite{peskin18,catalysis12,dzhioev11,preston21}, at the cost of a fully quantum description. 

A common accompanying assumption is the clear separation of time-scales between the classically described nuclei and the electronic environment, which through the use of a perturbative approximation allows for a Langevin description of the classical particle\cite{preston2020,preston21,prestonAC2020,kershaw20,Dou17,Dou2018,Bode12,pistolesi16,chen19,galperin19,Todorov12,fuse,nocera11}. In this regime, the particle dynamics are described by stochastic differential equations in which nuclei evolve under the influence of a frictional force which acts to subdue nuclear vibrations, and a stochastic force which delivers energy to the nuclei; the balance of these two forces yielding the {  effective} temperature of the molecule. In addition, the influence of the electronic environment on the molecular potential is captured by an additional, {  typically non-conservative\cite{dundas09,dundas12,cunningham14}}, renormalization term which can naturally describe switching phenomena and bond-stability\cite{preston2020,pistolesi16,Bode12,preston21}. The Langevin approach has had great success in the description and simulation of large-scale systems corresponding to real molecular electronic junction configurations, such as for graphene nanoribbons\cite{Brandbyge11,Christensen16}.

One such theoretically predicted phenomenon which leads to device breakdown is the notion of a negative dissipation; under a Langevin approach, this implies that the viscosity coefficient becomes negative\cite{preston2020,Bode12,fuse,Todorov12,hussein10,hopjan18,kartsev14}. In this regime, energy is applied to rather than dissipated from the classical coordinate until it reaches unsustainable temperatures for the device. Negative dissipations have also been predicted using purely quantum mechanical methods in which a population inversion in the quantized phonons leads to ever-increasing temperatures\cite{galperin19,hector18,Gunst13,Simine12,thoss11,Lu11,rizzi16}. The physicality of such theoretical results have often been called into question, thought to arise as an artifact of assumptions applied in the theory rather than a demonstration of real behaviour\cite{galperin18}.

In this paper, we extend the quasi-classical formalism by utilising a numerically exact method to calculate the forces acting on our classical coordinates without resorting to the assumption of a time-scale separation within the system. {  This is commonly known as the Ehrenfest approach to classical dynamics and it has garnered significant interest in the modelling of nuclear motion in nanoscale systems\cite{hyldgaard12,kartsev14,todorov10,Stefanucci06,Brandbyge03,peskin18,Metelmann11,dundas09,cunningham14,dundas12,hopjan18}. This approach treats the electron-nuclear interaction on a mean-field level; it therefore does not provide a full description of the inelastic scattering between electrons and nuclei and cannot fully capture the effects of joule heating within the system\cite{horsfield04,todorov07,Stefanucci06,horsfield2_04,todorov10}. Efforts have been made to remedy these caveats via perturbative corrections to the electron-nuclear correlations with some success\cite{todorov07,horsfield2_04}. However, in this study we seek only to utilise Ehrenfest dynamics as a basis for comparison for predictions made by the perturbative Langevin approach.} In doing so, we once again observe the emergence of negative dissipations and validate the results of the perturbative Langevin approximation, depending on the parameter range applied. 

Our approach additionally allows for the exploration of the autocorrelations in the stochastic force. {  In general, the stochastic force is correlated at different times dependent on the electronic structure of the system considered; this corresponds to coloured noise\cite{nitzanbook,chen19,todorov14}. The correlations are additionally dependent on the non-adiabatic motion of the classical coordinate. However, it is often computationally infeasible to account for coloured noise} and it becomes necessary to {  employ a white-noise approximation. Under this approximation, the coloured-noise diffusion is replaced by a Markovian, white-noise equivalent which attempts to produce the same dynamical behaviour\cite{preston2020,prestonAC2020,preston21,Bode12,Dou17,catalysis12}. }We apply our method to investigate the effects of a time-scale separation on the diffusion coefficient along with assessing the validity of the white-noise approximation to the diffusion under a variety of regimes.

In section \ref{Theory} we introduce the theory which describes the classical dynamics. This involves the introduction of the Ehrenfest force along with its corresponding perturbative Langevin approximation, both expressed in terms of non-equilibrium Green's functions. We also discuss our iterative method for evaluating the Ehrenfest force through time. In section \ref{singleDoF}, we demonstrate the accuracy of the perturbative approximation for a simple system consisting of a single classical degree of freedom, while in \ref{twoDoF} we then apply Ehrenfest dynamics to a system with two classical degrees of freedom. {  Finally, in section \ref{results_D} we utilise our method to evaluate the diffusion coefficient and the suitability of the} white-noise approximation for the stochastic force over a range of parameters. We use atomic units for all parameters and results presented.

\section{THEORY}
\label{Theory}
\subsection{Hamiltonian}
Our system is described by a generic tunnelling Hamiltonian as per
\begin{equation}
H(t) = {H}_{M}(t) + {H}_{L} + {H}_{R} + {H}_{LM} + {H}_{MR} + H_{\text{cl}}(t).
\label{hamiltonian}
\end{equation}
The total system Hamiltonian is partitioned into the following components; the molecular Hamiltonian ${H}_{M}(t)$, the left and right leads Hamiltonians ${H}_{L}$ and ${H}_{R}$, the leads-molecule coupling Hamiltonians ${H}_{LM}$ and ${H}_{MR}$ which describe the coupling between the electronic states on the central molecule and the left and right leads, respectively, and the classical Hamiltonian $H_{\text{cl}}$ which describes the time-evolving molecular geometry.

The molecular Hamiltonian takes the form: 
\begin{equation}
{H}_{M}(t) = \sum_{ij} h_{ij} ( \mathbf{x}(t) ) d^{\dag}_{i} d_{j},
\label{molecularhamiltonian}
\end{equation}
where the operators $d^{\dag}_{i}$ and $d_{j}$ denote the creation and annihilation operators for the {  atomic} electronic states whose energies are {  given by the diagonal Hamiltonian matrix elements $h_{ii}(\mathbf{x}(t))$ while the hopping amplitudes are given by $h_{ij}(\mathbf{x}(t))$}. Note the explicit time dependence here, which arises as a result of the time evolution of the multi-dimensional classical coordinate $\mathbf{x}$. We will usually not show the time dependence of $\mathbf{x}$ explicitly.

The leads Hamiltonian is taken in the standard form:
\begin{equation}
{H}_{L} + {H}_{R}  = \sum_{k \alpha} \epsilon_{k \alpha} d^{\dagger}_{k \alpha} d_{k \alpha},
\end{equation}
where the creation and annihilation operators are given by $d^{\dagger}_{k \alpha}$ and $d_{k \alpha}$, and the subscript $k\alpha$ denotes the operator acting on state $k$ in the $\alpha$ lead which has energy $\epsilon_{k \alpha}$.

The system-lead coupling Hamiltonians $H_{LM}$ and $H_{MR}$ are given by:
\begin{equation}
{H}_{LM} + {H}_{MR} = \sum_{k\alpha i} \Big( t_{k \alpha i}  d^{\dagger}_{k \alpha} d_{i} + \text{h.c.} \Big).
\end{equation}
The matrix elements $t_{k \alpha i}$ (and their conjugates) describe the tunnelling amplitudes between lead states $k \alpha$ and the molecular orbitals $i$. We take $t_{k \alpha i}$ to be independent of $\mathbf{x}$.

Finally, the classical Hamiltonian is given by
\begin{equation}
H_{\text{cl}}(t) = \sum_\nu \frac{p_\nu^2}{2m_\nu} + U(\mathbf{x}),
\end{equation}
where the summation is taken over all classical degrees of freedom. The classical momentum for the $\nu$ degree of freedom is given by $p_\nu$, while $m_\nu$ is the corresponding mass and $U (\mathbf{x})$ is the multi-dimensional potential.

\subsection{Forces on the Classical Coordinate}
{  The force operator for the force acting on the classical coordinates due to the quantum environment is given by
\begin{align}
\mathbf f(t)&=-\nabla H(t),
\\
&=- \nabla H_M(\mathbf{x}) - \nabla U(\mathbf{x}),
\label{heisenberg}
\end{align}
where $\nabla = [\partial_{\nu_1} ,\partial_{\nu_2} ,\hdots ]^T$ and $\partial_{\nu_1}$ is the partial derivative with respect to the classical coordinate $\nu_1$. We have retained only the Hamiltonian components which depend on the classical coordinates.} By taking the quantum average of the force operator, we introduce the so-called Ehrenfest force according to 
\begin{equation}
\mathbf F^{\text{ehr}}(t) = -\langle\nabla H_M (\mathbf{x})\rangle.
\end{equation}
This is the main object of the present study. As discussed, the Ehrenfest force alone is ill-equipped to capture the entirety of the dynamics of the system. In order to completely describe the classical motion, an additional stochastic force term is required according to 
\begin{equation}
f_{\nu}(t) = - \partial_\nu U + F^{\text{ehr}}_{\nu}(t) + \delta f_{\nu}(t),
\label{fulldynamics}
\end{equation}
where the subscript $\nu$ denotes the force acting on the $\nu$ classical degree of freedom. This equation can be obtained by taking the classical limit of the Feynman-Vernon influence functional for a system interacting with a quantum environment and applying a Hubbard-Stratonovich transformation, in which the dynamical evolution is split into a deterministic (Ehrenfest) component and a stochastic component\cite{Hu19,chen19}. For the purposes of our simulations, we disregard the stochastic component in (\ref{fulldynamics}) and instead focus solely on the Ehrenfest force along with the classical force arising from $U$ {  since an accurate description of joule heating is not the aim of this study.} We separately consider the influence of the stochastic force in Section \ref{results_D}. We choose to model the classical potential for any degree of freedom according to a harmonic potential as given by
\begin{equation}
U(\mathbf{x}) = \sum_\nu \frac{1}{2}k_{\nu} (\nu - \nu_0)^2,
\end{equation}
where $k_\nu$ is the spring constant for the $\nu$ degree of freedom, and $\nu_0$ is a parameter which shifts the potential along the $\nu$ coordinate. {  The model is is not restricted to this choice of potential; it is capable of dealing with arbitrary classical potentials.} Our primary tool for evaluating the forces is the non-equilibrium Green's function. For our Hamiltonian, the Ehrenfest force can be represented in terms of the lesser component of exact non-equilibrium Green's functions as {  
\begin{equation}
F_{\nu}^{\text{ehr}}(t) = i\text{Tr}\Big\{ \partial_{\nu} h(t) G^<(t,t)\Big\},
\label{ehrenfest_exact} 
\end{equation}}
{  where $h$ contains the elements of the molecular Hamiltonian and the elements of $G^<$ are defined according to 
\begin{equation}
G_{ij}^<(t,t') = i\langle d_j^\dag (t') d_i (t) \rangle.
\end{equation}
Each are square matrices with dimension corresponding to the number of molecular electronic energy levels. }

\subsection{Perturbative Expansion of the Ehrenfest Force}

By enforcing a time-scale separation between the classical degrees of freedom and the electronic environment, a perturbative expansion of the Ehrenfest force can be taken, from which we can derive the Langevin description of dynamics {\cite{preston2020,preston21,prestonAC2020,Bode12,Dou17,kershaw20,fuse,Todorov12}.} Here, our small parameter is $\Omega / \Gamma$, where $\Omega$ is the characteristic classical vibrational frequency and $1/ \Gamma$ is associated with the electron tunnelling time {  in molecular electronic junctions\cite{cuevasbook}.} This amounts to assuming that the classical coordinate vibrates over much longer time-scales than the time an electron spends in the central region. {  This interpretation is appropriate for the systems considered in our paper.} $\Gamma$ is generally a highly tunable experimental parameter while $\Omega$ depends on the vibrational mode considered. In order to identify the separate time-scales, we introduce Wigner coordinates for the time. These are the central time ($T$) and relative time ($\tau$) which are defined according to
\begin{equation}
T = \frac{t + t'}{2},
\;\;\;\;\;\;\;
\tau = t - t'.
\end{equation}
We then define an auxilliary force $\mathcal{F}(t,t')$ which is equal to $F^{\text{ehr}}(t)$ when $t=t'$. Thus, we have {  
\begin{equation}
\mathcal{F}(t,t')= i\text{Tr}\Big\{ \partial_{\nu} h(t) G^<(t,t')\Big\}.
\end{equation}}
Application of the Wigner transform along with its inverse results in  {  
\begin{equation}
\mathcal{F}(t,t')= i\text{Tr}\Big\{\int \frac{d\omega}{2\pi} e^{-i\omega \tau} \partial_{\nu} h(T) \widetilde{G}^<(T,\omega)\Big\}, 
\end{equation}}
where $G$ denotes a Green's function in the time domain, and $\widetilde{G}$ denotes the Wigner space  
defined as
\begin{equation}
\widetilde{G}(T,\omega) = \int d(t-t') e^{i \omega (t-t')} G(t,t').
\end{equation}

We adopt the convention that if the integral limits are not shown explicitly, they range from $-\infty$ to $\infty$. We will often subdue the $(T,\omega)$ indices for brevity. We now propose the ansatz,
\begin{equation}
\widetilde{G}^<=\widetilde{G}_{(0)}^< + \widetilde{G}_{(1)}^< + \widetilde{G}_{(2)}^<+ ...,
\label{ansatz}
\end{equation}
in which $\widetilde{G}_{(n)}^<$ is of $n^{th}$ order in our small parameter. {  Here, $\widetilde{G}_{(0)}^<$ is adiabatic and corresponds to the Born-Oppenheimer approximation, while the higher order terms go beyond this and account for the non-zero motion of the classical coordinates.} Applying this ansatz and letting $t=t'$ finally yields {  
\begin{align}
F^{\text{ehr}}_{\nu}(t)&= i\text{Tr}\Big\{\int \frac{d\omega}{2\pi}  \partial_{\nu} h(t) \Big(\widetilde{G}_{(0)}^< (t,\omega) + \widetilde{G}_{(1)}^< (t,\omega) + ...\Big) \Big\},
\\
&=\underbrace{F_{\nu,(0)} + F_{\nu,(1)}}_\text{retained in Langevin approach} + F_{\nu,(2)} + ....
\label{order_forces}
\end{align}}
Retaining only the zeroth (adiabatic) and first order contributions to the Ehrenfest force yields the deterministic forces present in the Langevin description, where the adiabatic force is purely position dependent while the first order force is the correction which accounts for non-zero velocities. The viscosity coefficient matrix $\hat{\xi}$ is obtained via
\begin{equation}
\mathbf{F_{(1)}} = -\hat{\xi} \mathbf{v},
\end{equation}
where $\hat{\xi}$ is an $n\times n$ matrix; $n$ corresponding to the number of classical degrees of freedom in the system.

\subsection{Solving for the Adiabatic and First Order Green's Functions}

The adiabatic and first order Green's functions in the frequency domain can be solved for via the Keldysh-Kadanoff-Baym equations, {  expressed in the Wigner space as \cite{Bode12,preston21,preston2020}}
{  
\begin{multline}
\Big(\omega+\frac{i}{2}\partial_{T}-e^{\frac{1}{2i}\partial_{\omega}^{G}\partial_{T}^{h}} h(T)\Big)\widetilde{G}^{R/A}=I\\
+e^{\frac{1}{2i}\left(\partial_T^\Sigma \partial_\omega^G-\partial_{\omega}^{\Sigma}\partial_{T}^{G}\right)}\widetilde{\Sigma}^{R/A}\widetilde{G}^{R/A},\label{eqm4}
\end{multline}
and 
\begin{multline}
\Big(\omega+\frac{i}{2}\partial_{T}-e^{\frac{1}{2i}\partial_{\omega}^{G}\partial_{T}^{h}} h(T)\Big)\widetilde{G}^{</>}=e^{\frac{1}{2i}\left(\partial_T^\Sigma \partial_\omega^G-\partial_{\omega}^{\Sigma}\partial_{T}^{G}\right)}\\
\times\Big(\widetilde{\Sigma}^{R}\widetilde{G}^{</>}+\widetilde{\Sigma}^{</>}\widetilde{G}^{A}\Big),\label{eqm3}
\end{multline}}
where we have shown the retarded/advanced and the lesser/greater collectively. {  Here, we adopt the convenient notation for derivatives, $\partial_T^G$, which denotes a partial derivative acting on the $G$ term with respect to $T$ and so on.} Again, each quantity here is a square matrix with dimensions corresponding to the number of electronic levels in the central region. {  Since $T$ is associated with the motion of the classical coordinate, terms containing $(\partial_T )^n$ will be of $n^\text{th}$ in our small parameter. Thus, we apply} our ansatz in (\ref{ansatz}), along with applying an expansion of the exponentials in orders of $\partial_T$. Truncating after the zeroth order and solving for $\widetilde{G}_{(0)}$ yields the standard adiabatic Green's functions as follows:
\begin{equation}
\widetilde{G}_{(0)}^{R/A} = \Big(\omega \hat{I}-h-\widetilde{\Sigma}^{R/A}\Big)^{-1}, 
\label{GRA}
\end{equation}
\begin{equation}
\widetilde{G}_{(0)}^{</>} = \widetilde{G}_{(0)}^R \widetilde{\Sigma}^{</>} \widetilde{G}_{(0)}^A.
\label{GLG}
\end{equation}
An equivalent approach for the first order yields\cite{Bode12,preston21,preston2020}
\begin{equation}
\widetilde{G}_{(1)}^{R/A} = \frac{1}{2i}\widetilde{G}_{(0)}^{R/A} \Big[\widetilde{G}_{(0)}^{R/A}, \partial_T h \Big] \widetilde{G}_{(0)}^{R/A},
\end{equation}
\begin{multline}
\widetilde{G}_{(1)}^{</>} = \widetilde{G}_{(0)}^R \widetilde{\Sigma}^{</>} \widetilde{G}_{(1)}^A + \widetilde{G}_{(1)}^R \widetilde{\Sigma}^{</>} \widetilde{G}_{(0)}^A 
\\
+ \frac{1}{2i} \widetilde{G}_{(0)}^R \Big(\partial_{T}h \widetilde{G}_{(0)}^R \partial_{\omega} \widetilde{\Sigma}^{</>} + \widetilde{G}_{(0)}^{</>} \partial_T h + h.c \Big)\widetilde{G}^A_{(0)}.
\end{multline}
The self-energy components in the frequency domain are given by
\begin{equation}
\widetilde{\Sigma}^R_{\alpha} = -\frac{i}{2}\Gamma_{\alpha},
\;\;\;\;\;\;
\widetilde{\Sigma}^A_{\alpha} = \frac{i}{2}\Gamma_{\alpha},
\label{sigmaAR}
\end{equation}
\begin{equation}
\widetilde{\Sigma}^<_{\alpha} = if_\alpha(\omega)\Gamma_{\alpha},
\;\;\;\;\;\;
\widetilde{\Sigma}^>_{\alpha} = -i[1-f_\alpha(\omega)]\Gamma_{\alpha}.
\label{sigma_lesser}
\end{equation}
Here, $f_{\alpha}(\omega)$ is the Fermi-Dirac distribution which describes the $\alpha$ lead. In solving for the above, we have applied the wide-band approximation to the leads such that $\Gamma_\alpha (\omega) = \Gamma_\alpha$. The explicit expression for an element of $\Gamma_{\alpha}$ is
\begin{equation}
\Gamma_{c,c',\alpha} = 2 \pi \rho_{\alpha} t^*_{c,\alpha}t_{c',\alpha},
\label{gamma}
\end{equation}
where $\rho_{\alpha}$ is the now constant density of states for the $\alpha$ lead, and $t_{c,\alpha}$ is the coupling strength between any state in the $\alpha$ lead and the central region electronic state $c$. We take $\Gamma_{\alpha}$ as a parameter for our model. In equations (\ref{sigmaAR}) to (\ref{gamma}), the full quantity can be obtained by taking a sum over the leads $\alpha$. For example, 
\begin{equation}
\Gamma_{c,c'} = \sum_{\alpha} \Gamma_{c,c',\alpha}.
\end{equation}

\subsection{Time-Stepping Approach to the Evolution of the Green's Function}
\label{algorithm}
Clearly, if we are able to calculate $G^<(t,t)$ at each time, then (\ref{ehrenfest_exact}) allows us to readily calculate the Ehrenfest force without need to resort to a perturbative approach involving time-scale separation. 
{Our method is based off of the seminal work of Jauho et al. in reference \cite{Jauho94}, from which we borrow the main defined quantities. A similar time-stepping solution to Ehrenfest dynamics was presented in reference \cite{Metelmann11}. However, our application of the approach as detailed in this section is more sophisticated and allows for better convergence and stability.}
In approaching these calculations, we borrow the main defined quantities from reference \cite{Jauho94}. The lesser and greater Green's function in the time domain are given by the Keldysh equation,
\begin{equation}
G^{</>} (t,t')=\int dt_1 \int dt_2 G^R(t,t_1)\Sigma^{</>} (t_1,t_2) G^A(t_2,t').
\label{keldysh}
\end{equation}
For the sake of clarity, in this section we will use $(...)$ to denote a functional dependence, whereas $[...]$ will denote a term in the equation. Now we will take (\ref{keldysh}) and express {  the $\Sigma^{</>}$ term} according to the inverse Wigner transform of (\ref{sigma_lesser}). In the lesser case, we find

\begin{widetext}

\begin{align}
G^< (t,t')&=i\sum_{\alpha} \int dt_1 \int dt_2 G^R(t,t_1) \int \frac{d\omega}{2\pi}e^{-i\omega [t_1 - t_2]}f_\alpha (\omega)\Gamma_\alpha G^A(t_2,t'),
\\
&=i\sum_{\alpha} \int dt_1 \int dt_2 G^R(t,t_1) \int \frac{d\omega}{2\pi}e^{-i\omega [t_1 - t_2]}f_\alpha (\omega)\Gamma_\alpha G^A(t_2,t)e^{i\omega[t-t]}e^{i\omega[t'-t']},
\\
&=i\sum_{\alpha} \int \frac{d\omega}{2\pi} f_\alpha (\omega) e^{i\omega [t' - t]} \int dt_1 e^{i\omega[t-t_1]} G^R(t,t_1) \Gamma_\alpha \int dt_2 e^{-i\omega [t' - t_2]} G^A(t_2,t'),
\\
&=i\sum_{\alpha} \int \frac{d\omega}{2\pi} e^{-i\omega \tau} f_\alpha (\omega) A(\omega,t) \Gamma_\alpha A^\dag(\omega,t').
\label{Glesser}
\end{align}
An equivalent derivation can be applied in the greater case to find

\begin{equation}
G^> (t,t')=-i\sum_{\alpha} \int \frac{d\omega}{2\pi} e^{-i\omega \tau} [1-f_\alpha (\omega)] A(\omega,t) \Gamma_\alpha A^\dag(\omega,t').
\label{Ggreater}
\end{equation}

\end{widetext}

Here we have defined the quantity,
\begin{equation}
A(\omega,t) = \int dt_1 e^{i\omega[t-t_1]} G^R(t,t_1).
\label{Adef}
\end{equation}
The evolution of $G^R$ is found via the Kadanoff-Baym equation in the time domain:
\begin{equation}
\Big(i\frac{\partial}{\partial t} - h(t)\Big)G^R (t,t') = \delta (t-t') + \int dt_1 \Sigma^R (t,t_1)G^R(t_1,t').
\label{KKB_time}
\end{equation}
By utilising the fact that 
\begin{equation}
\Sigma^R(t,t')=-\frac{i}{2}\Gamma\delta(t-t'),
\end{equation}
we can solve for the general solution of (\ref{KKB_time}) as
\begin{equation}
G^R(t,t') = -i\Theta(t-t')\hat{T}\text{exp}\Big\{-i\int_{t'}^t dt_1 h(t_1) - \frac{1}{2}\Gamma[t-t']\Big\}.
\label{GR_solution}
\end{equation}
Here, $\hat{T}$ is the time-ordering operator and $\Theta$ is the Heaviside step function. By substituting (\ref{GR_solution}) into (\ref{Adef}), we arrive at
\begin{multline}
A(\omega,t) =
\\
-i\int^t_{-\infty} dt_1 \hat{T}\text{exp}\Big\{i\omega[t-t_1] -i\int_{t_1}^t dt_2 h(t_2) - \frac{1}{2}\Gamma[t-t_1] \Big\}.
\label{A}
\end{multline}
This is the quantity which we will iteratively evolve forwards in time, from which we can then extract $G^<$ via (\ref{Glesser}). We begin by applying a step $\Delta t$ forwards in time to (\ref{A})),

\begin{widetext}

\begin{align}
A(\omega,t+\Delta t) &= -i\int^{t+\Delta t}_{-\infty} dt_1 \hat{T}\text{exp}\Big\{i\omega[t+\Delta t-t_1] -i\int_{t_1}^{t+\Delta t} dt_2 h(t_2) - \frac{1}{2}\Gamma[t+\Delta t-t_1] \Big\},
\\
&=-i\int^{t}_{-\infty} dt_1 \hat{T}\text{exp}\Big\{i\omega[t+\Delta t-t_1] -i\int_{t_1}^{t+\Delta t} dt_2 h(t_2) - \frac{1}{2}\Gamma[t+\Delta t-t_1] \Big\}
\\
& -i\int^{t+\Delta t}_{t} dt_1 \hat{T}\text{exp}\Big\{i\omega[t+\Delta t-t_1] -i\int_{t_1}^{t+\Delta t} dt_2 h(t_2) - \frac{1}{2}\Gamma[t+\Delta t-t_1] \Big\},
\label{AB}
\\
&= A^A (\omega , t+\Delta t) + A^B (\omega , t + \Delta t).
\label{AA_and_AB}
\end{align}

Here we have split $A$ into $A^A$ which reflects how the history before $t$ informs the system at $t+\Delta t$, and $A^B$ which contains the effects of the system from $t$ to $t + \Delta t$. $A^A$ can be further simplified by partitioning the inner integral as follows:

\begin{align}
A^A(\omega,t+\Delta t) &= -i\int^{t}_{-\infty} dt_1 \hat{T}\text{exp}\Big\{i\omega\Delta t -i\int_{t}^{t+\Delta t} dt_2 h(t_2) - \frac{1}{2}\Gamma\Delta t \Big\}
\\
&\times \hat{T}\text{exp}\Big\{i\omega[t-t_1] -i\int_{t_1}^{t} dt_3 h(t_3) - \frac{1}{2}\Gamma[t-t_1] \Big\},
\\
&=\hat{T}\text{exp}\Big\{i\omega\Delta t -i\int_{t}^{t+\Delta t} dt_2 h(t_2) - \frac{1}{2}\Gamma\Delta t \Big\} A(\omega,t).
\label{AA}
\end{align}

\end{widetext}

We will make the assumption that for a sufficiently small $\Delta t$, $h(t)$ can be approximated as being piecewise constant. Thus, at some time-step $t_n$, we assume that for $t_n < t < t_{n+1}$, we have $h(t)=h(t_n)$. This entails that our time step size should be sufficiently small in the computational implementation when calculating classical trajectories. This enables us to directly calculate the integrals over $h(t)$ appearing in (\ref{AA}) and (\ref{AB}) along with removing the time-ordering operators. We thus re-express $A^A(\omega,t+\Delta t)$ and $A^B(\omega,t+\Delta t)$ according to
\begin{equation}
A^A(\omega,t+\Delta t) = \text{exp}\Big\{i\Delta t\Big(\omega -h(t) + \frac{i}{2}\Gamma \Big) \Big\} A(\omega,t), 
\label{AA_final}
\end{equation}
and 
\begin{align}
A^B(\omega,t+\Delta t) &= -i\int^{\Delta t}_{0} dt_2 \text{exp}\Big\{i t_2 \Big(\omega -h(t) + \frac{i}{2}\Gamma \Big) \Big\},
\\
&= -i\Big(e^{\Lambda\Delta t} - I\Big)\Lambda^{-1},
\label{AB_final}
\end{align}
where $I$ is the identity matrix and for cleanliness, we have used $\Lambda = i\Big(\omega-h(t)+\frac{i}{2}\Gamma \Big)$. Here we have used the matrix exponential identity:
\begin{equation}
\int^{T}_0 dt e^{X t} = \Big( e^{X T} - I\Big)X^{-1}.
\end{equation}
Now we have everything we need to be able to perform an Ehrenfest simulation of the classical dynamics. The algorithm functions as follows. First, we assume an adiabatic initial condition, such that (\ref{Adef}) can be calculated through the use of (\ref{GRA}). The initial force can then be computed according to (\ref{Glesser}) and (\ref{ehrenfest_exact}) given the chosen initial conditions for the classical coordinate. The calculated Ehrenfest force feeds into an external algorithm which updates the nuclear dynamics to the next time step, at which point we use (\ref{AA_final}) and (\ref{AB_final}) to non-adiabatically compute $A$ at the next time step. From here the process repeats, yielding a wholly non-adiabatic, classical trajectory in time.

\section{RESULTS}
\label{Results}

\subsection{Purely Positive Viscosity in the Single-Level Case}

In the case where there is only a single electronic level in the central region and only the molecular Hamiltonian contains classical dependence, the diagonal components of the viscosity matrix $\xi$ can be given by
\begin{equation}
\xi_{\nu \nu} = \frac{\Gamma(\partial_{\nu}h)^2}{4\pi}
\int d\omega \frac{\frac{f_L}{T_L}(1-f_L)\Gamma_L 
+ \frac{f_R}{T_R}(1-f_R)\Gamma_R}{\Big((\omega-h)^2+\Gamma^2/4\Big)^2},
\label{singleLevelVisc}
\end{equation}
{  where $T_L$ and $T_R$ are the temperatures for the left and right lead, respectively. In our model, we set $T_L=T_R = 0.001$.} A rigorous derivation of (\ref{singleLevelVisc}) has been relegated to Appendix \ref{appendix_xi}. Thus, we observe that the diagonal viscosity elements are purely positive as each term in (\ref{singleLevelVisc}) is positive. From a Langevin standpoint, theory is known to produce negative viscosities in non-equilibrium in the case of multiple electronic levels\cite{Bode12,hussein10,Todorov12}, as well as resulting from non-Condon effects in which the geometry at the leads-interface becomes relevant\cite{preston2020}. Due to the cumbersome nature of the equations in the latter case, we focus only on the former in this paper.

\subsection{Single Classical Degree of Freedom}
\label{singleDoF}

\begin{figure}
\includegraphics[width=0.5\textwidth]{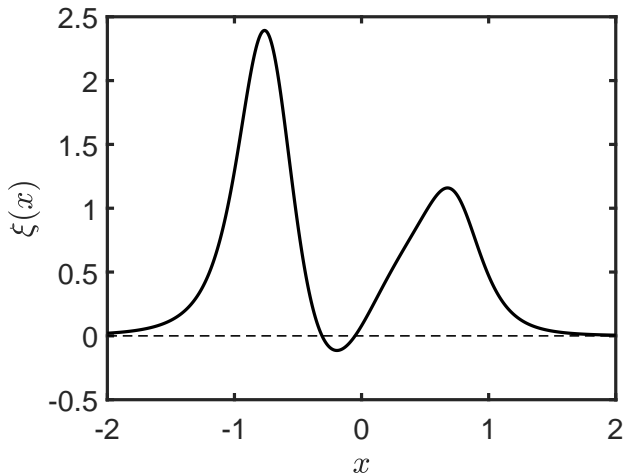}
\caption{The viscosity coefficient as a function of the classical coordinate. {  Atomic units are used for quantities here and in all further plots in the paper.}}
\label{neg}
\end{figure}

Here we present results for a two-level system with a single classical degree of freedom. The matrix elements of the molecular Hamiltonian take the form
\begin{equation}
H_M = 
\begin{pmatrix}
-\lambda x & \eta \\
\eta & \lambda x
\end{pmatrix}
,
\end{equation}
where the classical coordinate $x$ has the effect of linearly shifting the two atomic orbitals and $\eta$ dictates the overlap. We choose the left lead to be coupled to the first state and the right lead to be coupled to the second state which gives our $\Gamma$ matrices the following form:
\begin{equation}
\Gamma_{L} = 
\begin{pmatrix}
\gamma & 0 \\
0 & 0
\end{pmatrix},
\;\;\;\;\;\;
\Gamma_{R} = 
\begin{pmatrix}
0 & 0 \\
0 & \gamma
\end{pmatrix}.
\end{equation}
In this section, the common parameters are $\lambda = \eta = 0.1$, $\gamma = 0.05$, $k = 1$, $x_0 = -0.22$ and $\mu_L =- \mu_R = 0.12$, where $\mu_\alpha$ is the chemical potential of the $\alpha$ lead. In contrast to the single-level case, this model allows us to observe negative viscosities. An example of this is shown in Figure \ref{neg} where the coordinate dependent viscosity coefficient {  (now a scalar function in the case of a single classical degree of freedom)} becomes negative in a small region around $x\approx -0.2$.

\begin{figure}
\includegraphics[width=0.5\textwidth]{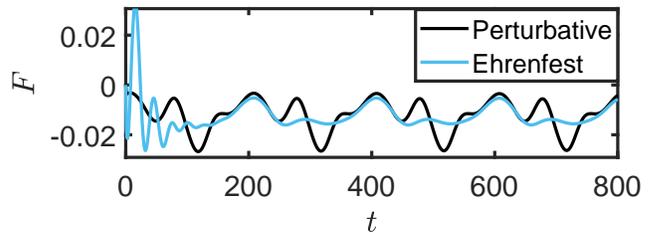}
\caption{Trajectory in time of $F^{\text{ehr}}(t)$ (Ehrenfest) and $F_{(0)}(t) + F_{(1)}(t)$ (perturbative). $\Omega /\Gamma\approx 0.32$.}
\label{Forces}
\end{figure}

\begin{figure}
\centering
\includegraphics[width=0.5\textwidth]{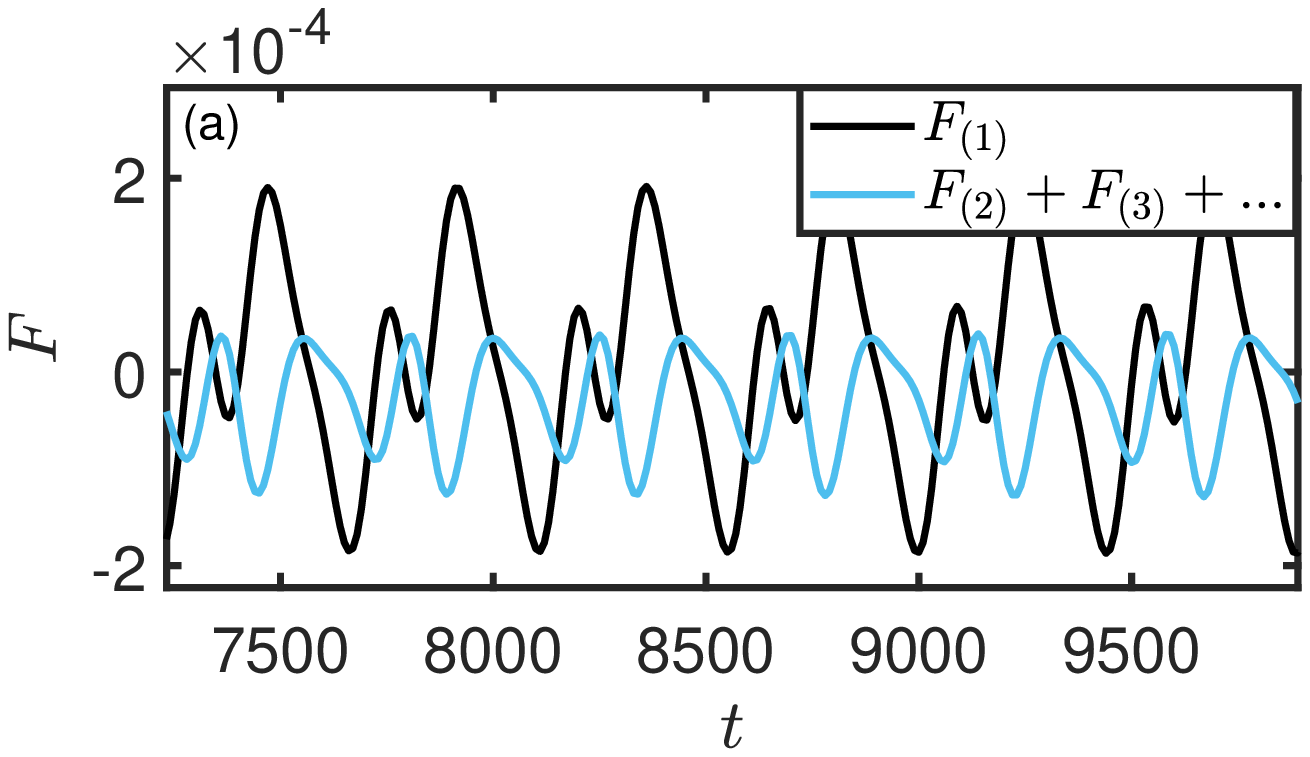}
\includegraphics[width=0.5\textwidth]{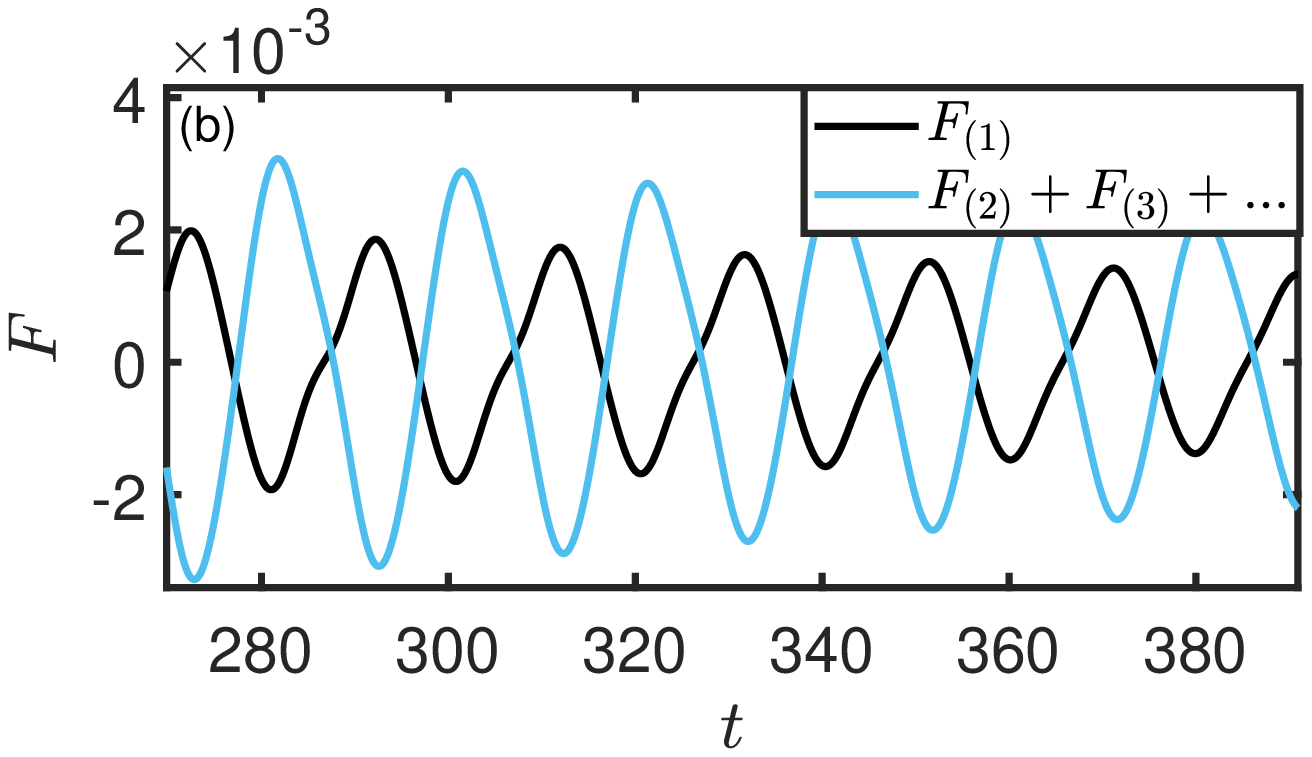}
\caption{Comparison of the first order force against the sum of all higher order forces. (a) $\Omega /\Gamma\approx 0.14$, (b) $\Omega /\Gamma\approx 3.2$.}
\label{higherOrder}
\end{figure}

By utilising our time-stepping algorithm to evolve $G^<$ in time as presented in \ref{algorithm} we can calculate the Ehrenfest force as a function of time, from which we can then computationally simulate the dynamics of the classical coordinate. In Figure \ref{Forces}, we calculate a classical Ehrenfest trajectory and record $F^{\text{ehr}}$ at each time step. We also record the corresponding forces which the classical coordinate would experience under the perturbative truncation of $F^{\text{ehr}}$ after the first order. We note that the methods clearly differ in the transient regime depending on the chosen initial conditions, before each settles into periodicity as the coordinate oscillates in time. The cause of the differences between the calculated Ehrenfest and perturbative forces are made clearer in Figure \ref{higherOrder}, where a direct comparison has been made of the magnitude of the first order force relative to the sum of all higher order forces ($F_{(2)} + F_{(3)} + ...$) for different values of $\Omega / \Gamma$. This is achieved by varying the effective mass of the classical coordinate while keeping all other parameters constant. We can then estimate $\Omega$ by assuming that the oscillations are approximately harmonic. We observe that when $\Omega / \Gamma$ is small in (a), the higher order forces have only a small relative contribution. In (b) however, $\Omega / \Gamma$ is no longer small and thus our perturbative assumption is no longer satisfied, meaning the higher order forces have become increasingly relevant.

\begin{figure}
\centering
\includegraphics[width=0.5\textwidth]{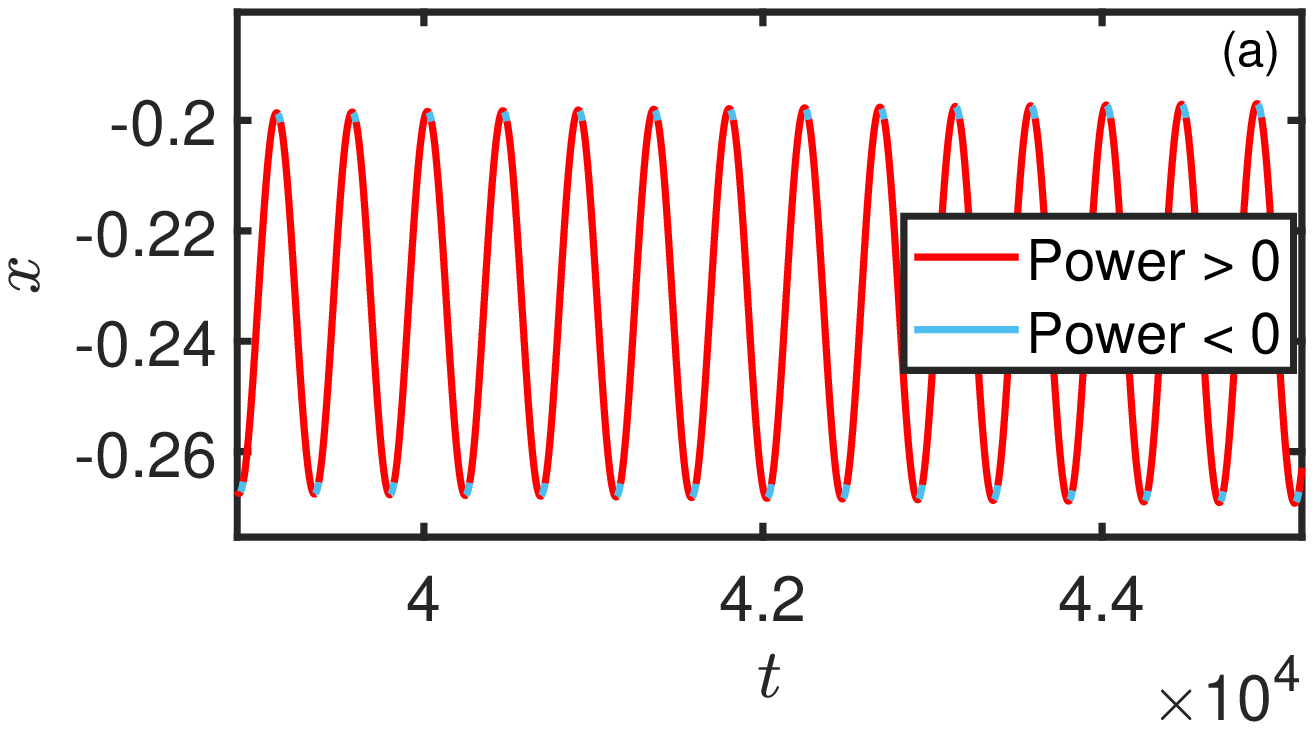}
\includegraphics[width=0.5\textwidth]{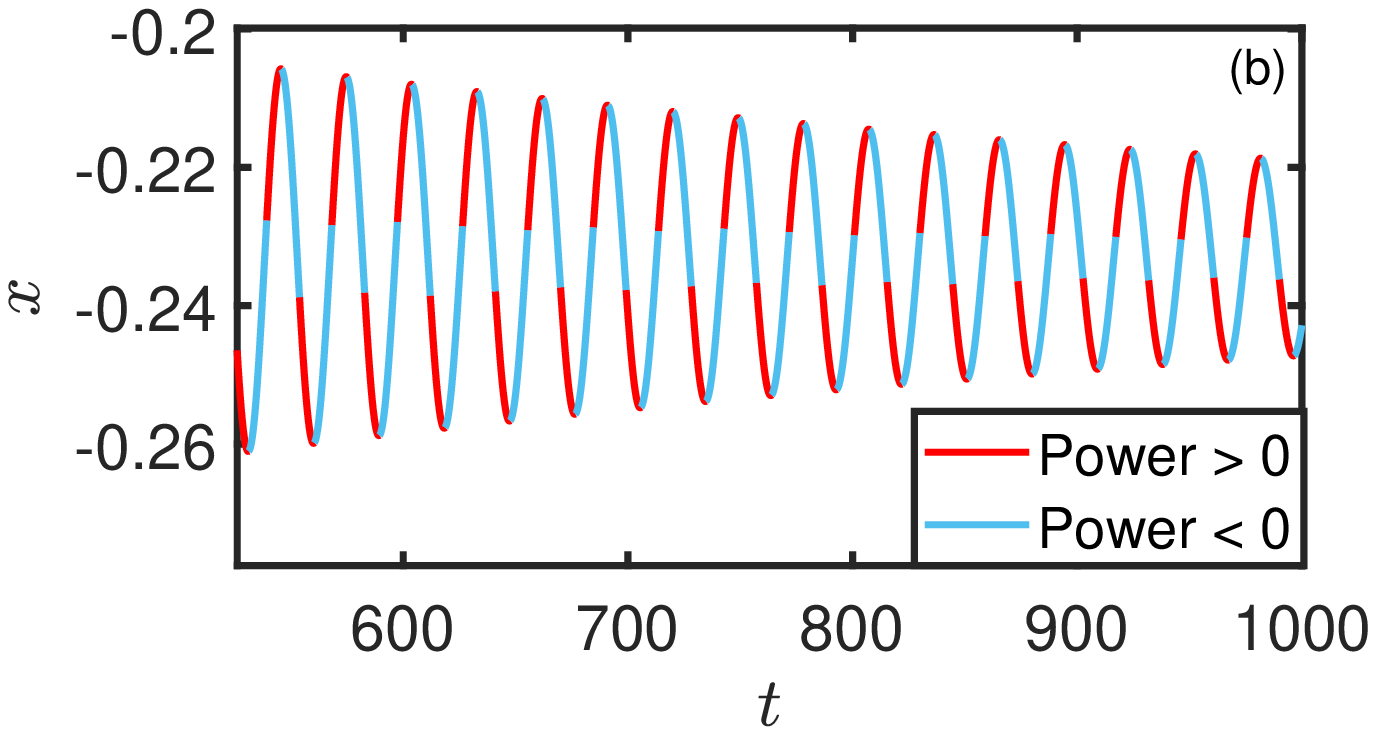}
\includegraphics[width=0.5\textwidth]{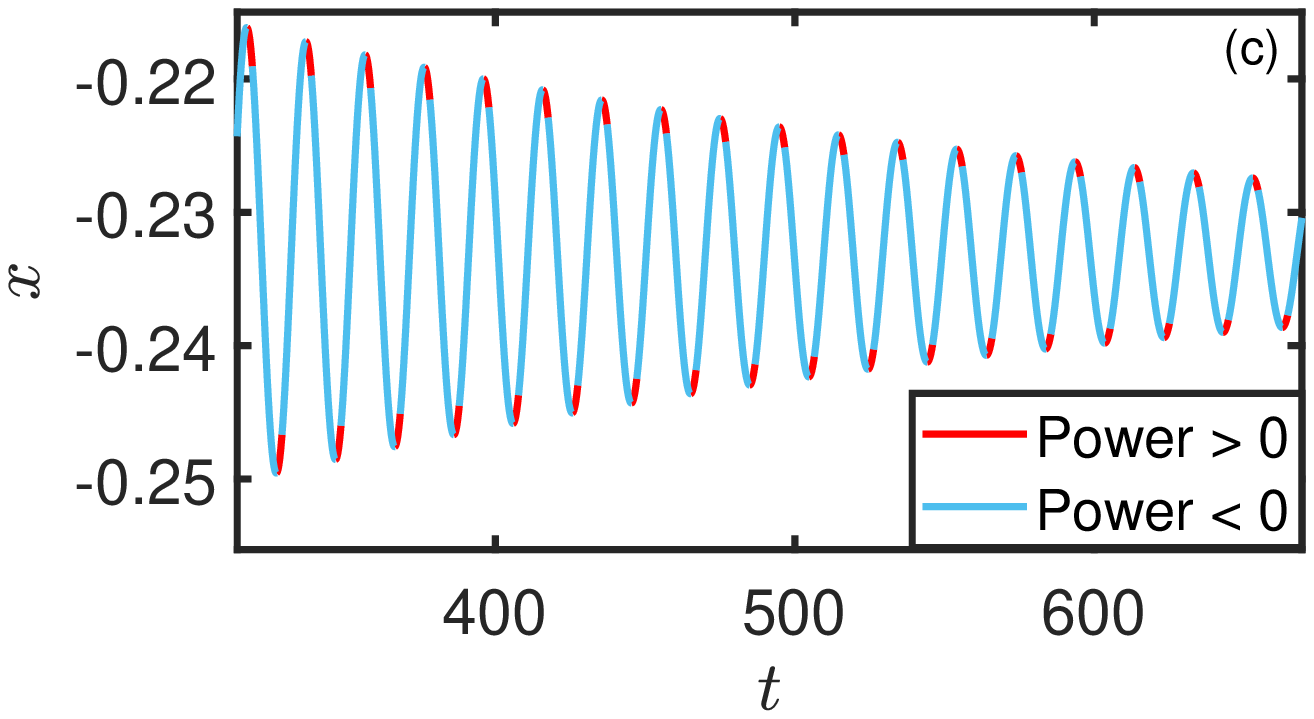}
\caption{Samples of Ehrenfest trajectories of the classical coordinate against time for (a) $\Omega /\Gamma\approx 0.14$, (b) $\Omega /\Gamma\approx 2.2$, (c) $\Omega /\Gamma\approx 3.2$. Colour shows the instantaneous power of the classical coordinate.}
\label{trajectories}
\end{figure}

\begin{figure}
\includegraphics[width=0.5\textwidth]{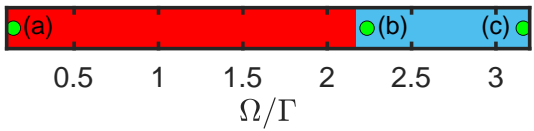}
\caption{Regimes of negative (blue) or positive (red) average power input to the classical coordinate over a period of oscillation. The green points refer to the corresponding plots in Figure \ref{trajectories}.}
\label{regimes}
\end{figure}

\begin{figure}
\centering
\includegraphics[width=0.5\textwidth]{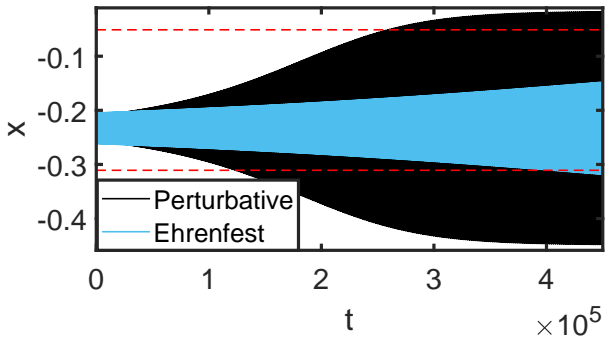}
\caption{{  Comparison of long trajectories of the classical coordinate in time using the Ehrenfest approach and the truncated perturbative approach. The viscosity coefficient is negative between the red dashed lines and positive outside. $\Omega /\Gamma \approx 0.14$.} }
\label{long_trajectories}
\end{figure}

Calculation of the time-dependent forces allows us to then simulate the phase-space trajectory of the classical coordinate as a function of time. Figure \ref{trajectories} shows time dependent trajectories of the classical coordinate $x$ for three different values of the small parameter $\Omega / \Gamma$. We also calculate the instantaneous power of the classical coordinate {  at each time step according to $(F^{\text{ehr}} - F_{(0)})\cdot \frac{p}{m}$, which includes only the effects of the excitational and dissipative forces.} The classical coordinate is intentionally confined to the region of negative viscosity shown in Figure \ref{neg} via our choice of $U(x)$ such that under the perturbative assumption, the instantaneous power will be positive at all time-steps. We observe that when $\Omega / \Gamma \ll 1$ as in (a); rendering the perturbative approximation as valid, the instantaneous power is overwhelmingly positive in agreeance with the perturbative approximation. This entails that the oscillations will increase until reaching regions of positive viscosity, whereby they will form a limit cycle. {  This is further illustrated in Figure \ref{long_trajectories} via long trajectories for the same parameters, where the amplitude of oscillations increases in both the Ehrenfest and perturbative approaches. This result demonstrates that even in the absence of time-scale separation, negative viscosities emerge and dictate the behaviour of the system. However, we observe the resultant trajectories from the two methods to diverge from each other over longer time-scales, demonstrating the importance of the higher order forces even for these parameters where we expect they can be reasonably neglected. We note that Ehrenfest dynamics has been used to observe the effects of negative viscosities in different regimes in the literature\cite{Metelmann11,hopjan18,kartsev14,rizzi16}.} In Figure \ref{trajectories} (b) and (c) where $\Omega / \Gamma > 1$ and our perturbative assumption is no longer adequate, we observe that instantaneous power is more often negative wherein the electronic environment is taking energy away from the classical coordinate. This means that the higher order terms (2nd order and above) in our perturbative expansion for $F^{\text{ehr}}$ have become more relevant and the dissipative nature of these forces are overwhelming the negative viscosity produced by $F_{(1)}$. These results are summarised in Figure \ref{regimes} in which the average power input to the classical coordinate over a period of oscillation is calculated and classified as positive (blue) or negative (red). We observe that the average power input to the classical coordinate is positive far beyond when the perturbative assumption is valid. We anticipate that this cut-off between positive and negative average power input is highly dependent on the model and parameters at hand. However, this demonstrates that while negative viscosities will still emerge under a numerically exact approach, it can be dominated by dissipative higher order forces for large values of $\Omega / \Gamma$. {  These results suggest that the inclusion of the higher order forces which emerge through Ehrenfest dynamics acts to further subdue classical vibrations within the system.} 

\subsection{Two Classical Degrees of Freedom}
\label{twoDoF}

The algorithm for calculating $F^{\text{ehr}}$ can be readily extended to account for many classical degrees of freedom. In this section, we consider another 2-level model consisting now of two classical degrees of freedom; a stretching component and an angular component. {As a result, in contrast to the case of a single vibrational degree of freedom, $F_{(0)}$ now becomes typically non-conservative.}
The stretching coordinate $q$ modulates the hopping amplitude between electronic sites in the central region, while the angular coordinate $\theta$ accounts for the shift in electric levels due to the electric field produced by the applied voltage bias. Thus, the molecular Hamiltonian now takes the form
\begin{equation}
H_M = 
\begin{pmatrix}
h_L(\theta) & \eta(q) \\
\eta(q) & h_R(\theta)
\end{pmatrix}
.
\end{equation}
The hopping amplitude is given by 
\begin{equation}
\eta(q)=\eta e^{-q}\Big(1+q+q^2/3\Big).
\end{equation}
This is a generic $1s$-orbital overlap scaled by some constant $\eta$\cite{mcquarrie}. Meanwhile, the atomic levels are given by
\begin{equation}
h_{L/R}(\theta) = \pm E q_0 \cos (\theta),
\end{equation}
where $q_0$ is the constant bond-length parameter which appears in our classical potential for $q$, while $E$ is the linear approximation to the electric field across the junction as determined by
\begin{equation}
E=\frac{\mu_L - \mu_R}{L},
\end{equation}
and $L$ is the junction length. For all results in this section, we set $\eta = 0.1$, $q_0 = 2$, $\theta_0 = 0$, $L = 4$, $k_q = 0.1$, $k_\theta = 0.05$,  and $\mu_L = -\mu_R = 0.1$. All other components of the full Hamiltonian take the same form as in section \ref{singleDoF}.

\begin{figure}
\centering
\includegraphics[width=0.5\textwidth]{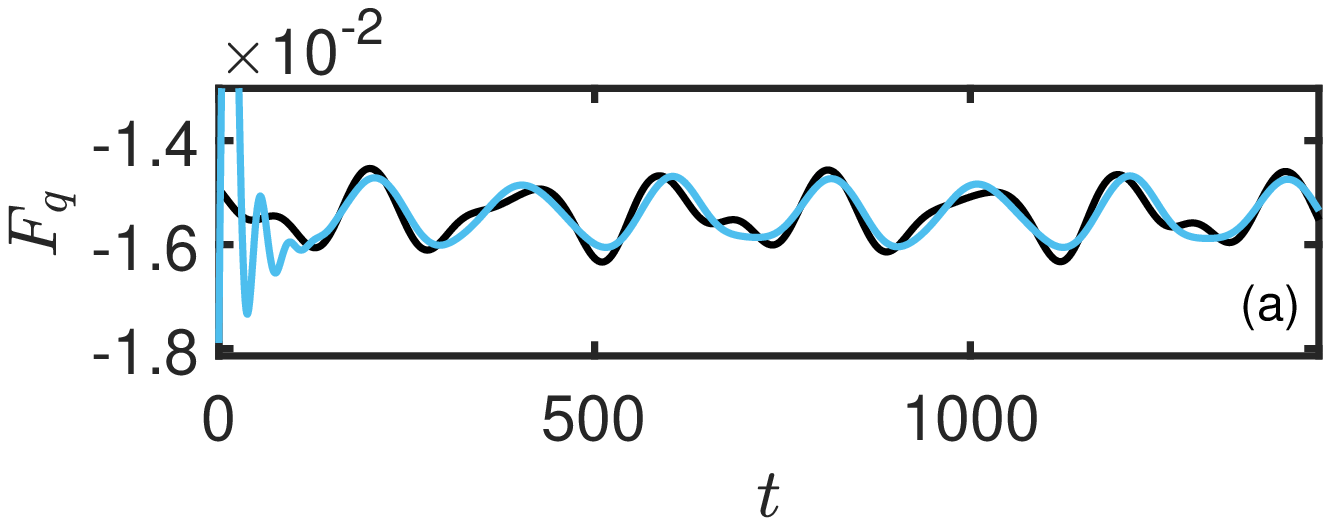}
\hspace*{0.25cm}\includegraphics[width=0.48\textwidth]{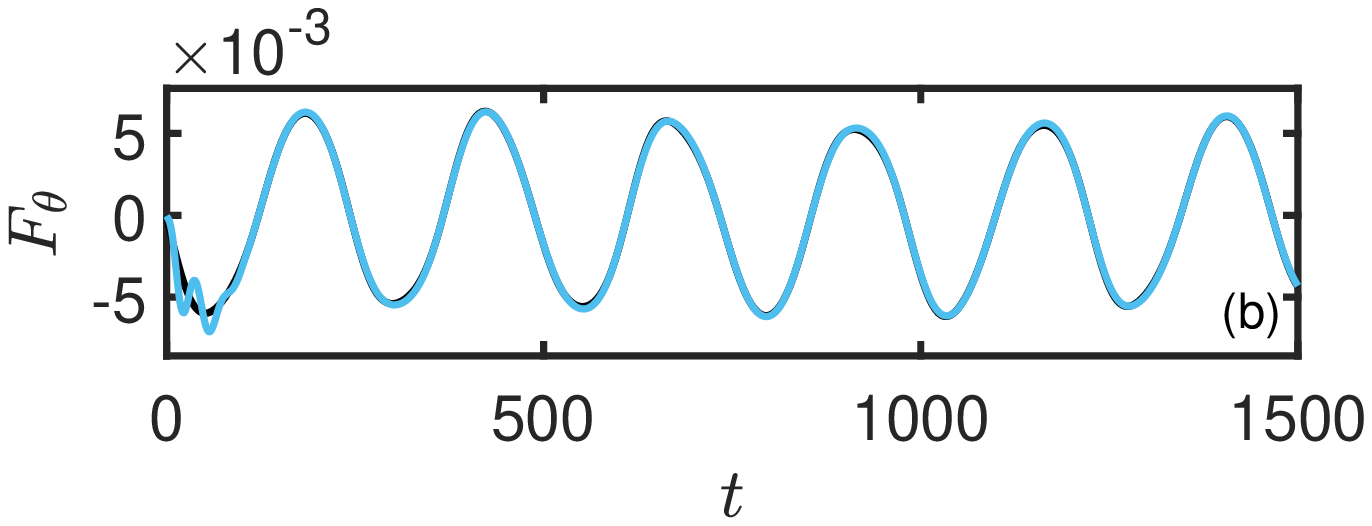}
\caption{The force trajectories in time for the Ehrenfest force (blue) and perturbative force (black) acting on (a) $q$ and (b) $\theta$. $\Omega_q / \Gamma \approx 0.32$, $\Omega_\theta / \Gamma \approx 0.22$.}
\label{2dof_forces}
\end{figure}

\begin{figure}
\centering
\includegraphics[width=0.5\textwidth]{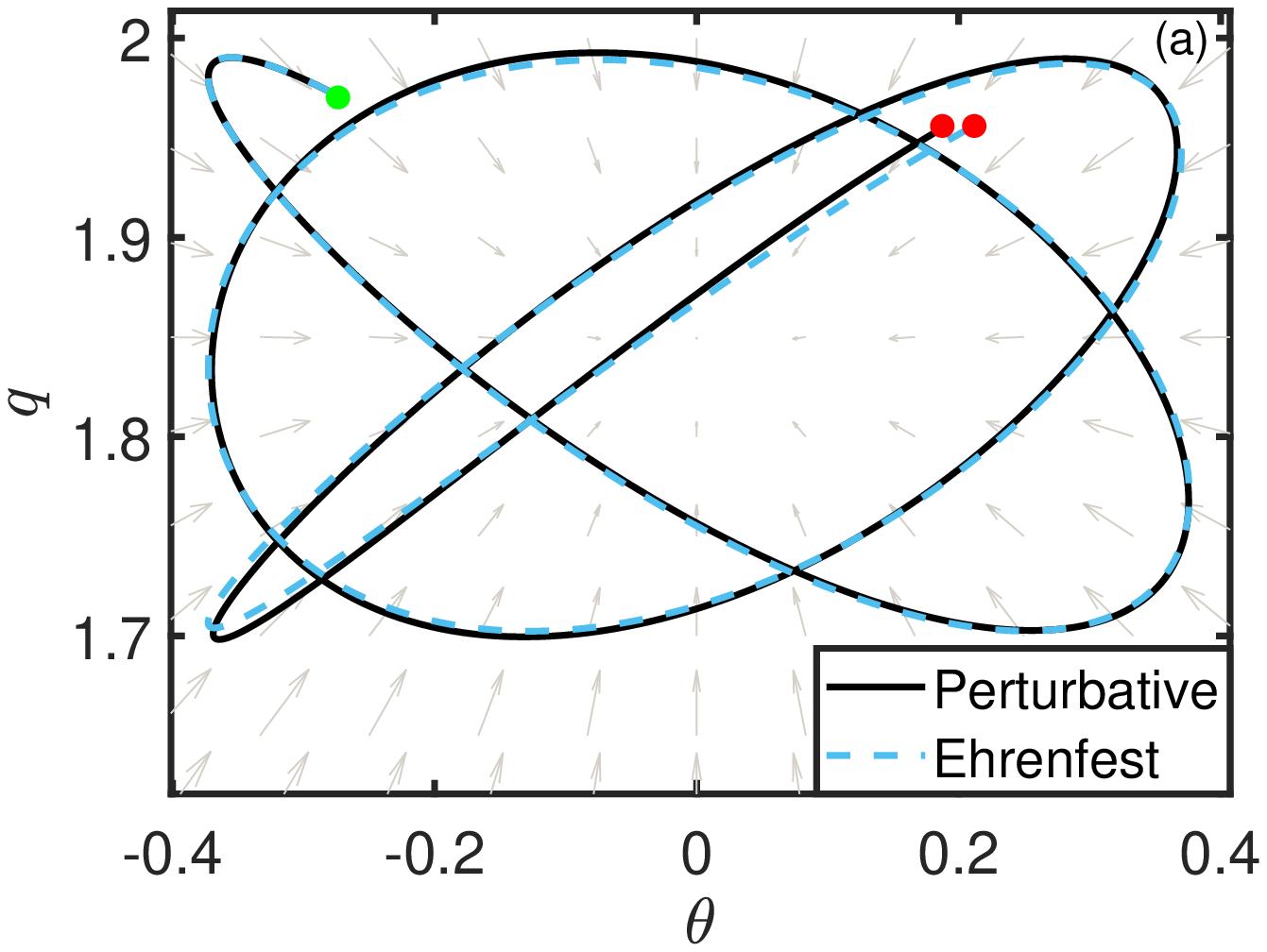}
\includegraphics[width=0.5\textwidth]{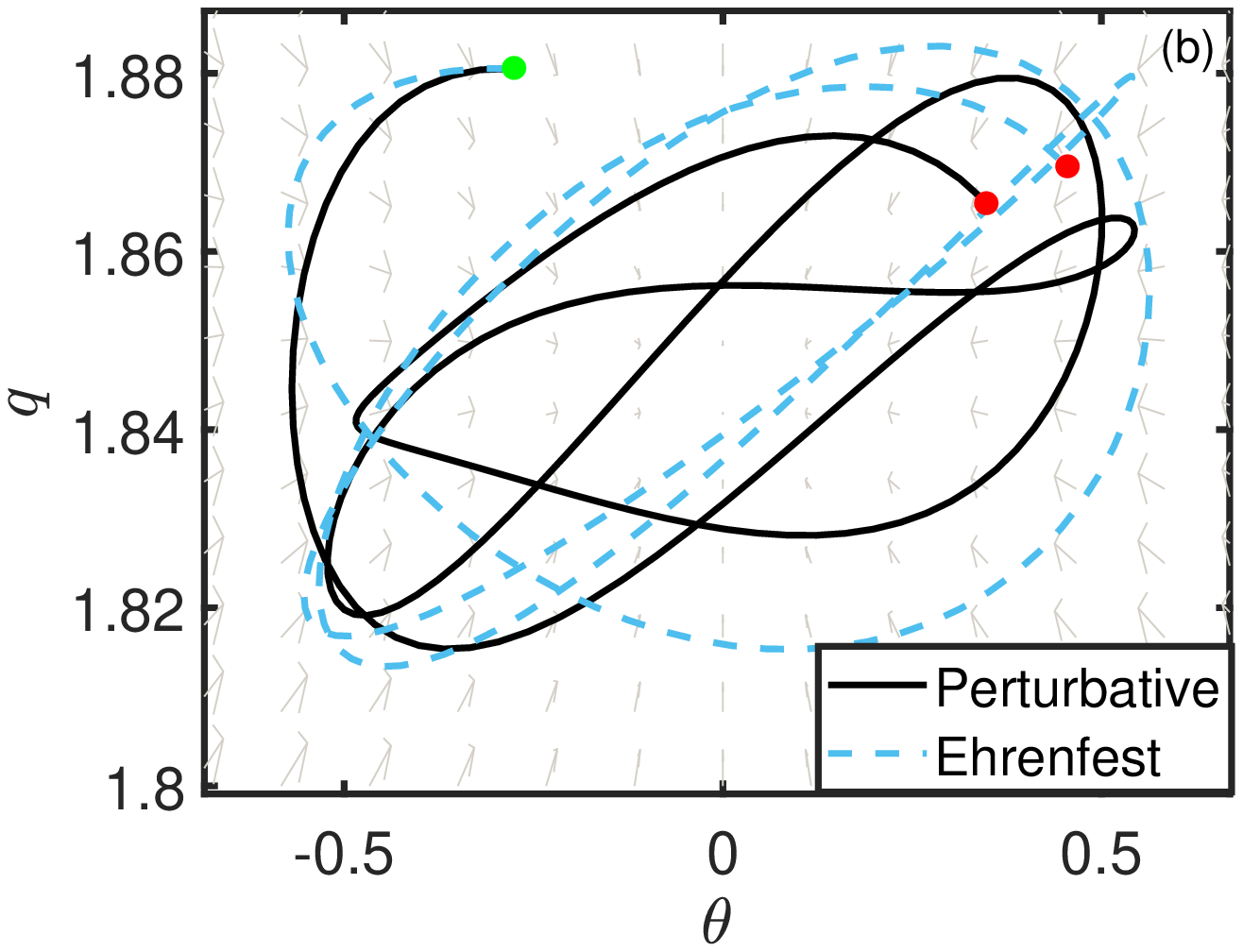}
\caption{Short classical trajectories in the coordinate-space comparing the Ehrenfest method and perturbative method. Green(red) points denote the start(end) of the trajectory. The vector field shows $F^{cl}+F_{(0)}$. (a) $\Omega_q / \Gamma \approx 0.32$, $\Omega_\theta / \Gamma \approx 0.22$, and a trajectory length of $600$, and (b) $\Omega_q / \Gamma = 1$, $\Omega_\theta / \Gamma \approx 0.7$, and a trajectory length of $200$.}
\label{2dof_traj}
\end{figure}

In Figure \ref{2dof_forces}, we calculate an Ehrenfest trajectory and record the Ehrenfest and perturbative forces on each degree of freedom separately. Given that we have more than one classical degree of freedom, this also includes the so-called non-equilibrium anti-symmetric forces in which the motion of $\theta$ induces a force on $q$ and vice versa; the net anti-symmetric force being perpendicular to the motion of the classical coordinate.
Figure \ref{2dof_traj} performs a direct comparison between the time dependent trajectories calculated via the Ehrenfest force and the perturbative approximation. To do so, we first calculate a classical Ehrenfest trajectory of our two coordinates. From this trajectory at some point in time after the unusual transient behaviour in $F^{\text{ehr}}$ has subsided, we extract the initial conditions for a comparative trajectory where the forces are calculated according to the perturbative approximation. In Figure \ref{2dof_traj} (a), the trajectories are compared for large effective masses for each coordinate such that the perturbative assumption should be satisfied. We observe only small differences between the trajectories resulting from the different methods in this case. Contrarily, Figure \ref{2dof_traj} (b) demonstrates for small effective masses that the trajectories deviate away from each other almost instantly and undergo largely different oscillatory behaviour. This is because the perturbative assumption is no longer valid since $\Omega / \Gamma$ is not sufficiently small.

{  
\subsection{Evaluating the Diffusion Coefficient}
\label{results_D}

The stochastic force term which we have thus far avoided in (\ref{fulldynamics}) is a stochastic Gaussian process entirely defined by the following\cite{chen19}:
\begin{equation}
\langle \delta f (t) \rangle = 0,
\;\;\;\;\;\;\
\langle \delta f (t) \delta f (t') \rangle = D(t,t'),
\label{D}
\end{equation}
where we will refer to $D(t,t')$ as the exact diffusion coefficient. We will consider a system consisting of a single electronic level coupled to a single classical degree of freedom. Our molecular Hamiltonian is given by
\begin{equation}
H_M = h_0 + \lambda x,
\end{equation}
where $x$ represents any generic classical degree of freedom. We will let $h_0 = 0$ and our $\Gamma_{\alpha}$ become scalar inputs in the single level case. The exact diffusion coefficient in (\ref{D}) for our model can then be expressed in terms of nonequilibrium Green's functions according to 
\begin{equation}
D(t,t') = \lambda^2 G^> (t,t') G^< (t',t).
\label{our_D}
\end{equation}
A derivation of this result is presented in Appendix \ref{appendix_D}. Our $D(t,t')$ here is not exclusively real. Since our stochastic force is classical, {  we symmetrize (\ref{our_D}) by taking the real component such that our expression for $D(t,t')$ is now in correspondence with reference \cite{chen19} as per
\begin{equation}
D(t,t') = \text{Re}\left\{\lambda^2 G^> (t,t') G^< (t',t)\right\}.
\end{equation}}
$D(t,t')$ accounts for the effects of the random fluctuations about the mean-field of the electronic environment on the classical degrees of freedom along with accounting for the feedback of the classical coordinate on the electronic environment due to its motion. However, generally a time-scale separation within the system is utilised in order to produce a perturbative solution to $D(t,t')$ in which the non-adiabatic feedback due to the motion of the classical coordinate is not included; in other words, the classical coordinate evolves adiabatically. We can solve for the adiabatic form of $D(t,t')$ by considering the Wigner transform of (\ref{our_D}) and retaining only the adiabatic terms. With some work which has been relegated to Appendix \ref{appendix_col}, the Wigner transform of (\ref{our_D}) can be expressed as
\begin{equation}
\tilde{D}(\omega,T) = \lambda^2 \int \frac{d\omega '}{2\pi} \tilde{G}^> (\omega + \omega ', T)\tilde{G}^< (\omega ',T).
\label{D_exact_wig}
\end{equation}
{  We note that (\ref{D_exact_wig}) is a result of the phonon self-energy derived in reference \cite{Todorov12}.} Now, the adiabatic diffusion is found by taking $\tilde{G}^{</>}$ to be our adiabatic Green's functions as in (\ref{GLG}). The subsequent application of the inverse Wigner transform then yields the adiabatic diffusion in the time domain as 
\begin{equation}
D_{(0)}(\tau,T) = \lambda^2 \int \frac{d\omega}{2\pi} e^{-i\omega \tau} \int \frac{d\omega '}{2\pi} \tilde{G}_{(0)}^> (\omega + \omega ', T)\tilde{G}_{(0)}^< (\omega ',T).
\label{D_ad}
\end{equation}
The adiabatic diffusion will serve as a base of comparison with the exact diffusion in order to analyse the effects of a time-scale separation within the system. We additionally analyse the validity of the white-noise approximation for different parameters. The white-noise approximation is a method of coarse-graining the diffusion coefficient which allows for a simpler mathematical and computational treatment. This is done by assuming that the coloured-noise exact diffusion coefficient can instead be replaced by a white-noise equivalent according to 
\begin{equation}
D(t,t') \equiv D^w (T) \delta(t-t'),
\label{Dw_orig_def}
\end{equation}
where we have introduced the white-noise diffusion coefficient $D^w$. Thus, the aim is to accurately reproduce the effects of the stochastic force under the assumption that it is entirely uncorrelated in time. In order to replicate the correct dynamics using this approximation, it can be shown that $D^w$ must then take the form {  
\begin{equation}
D^w (t_f)= \int^{t_f}_{-t_f} d\tau D(t_f,\tau),
\label{Dw_integral}
\end{equation}
where to avoid ambiguity, $t_f$ denotes a specific point along the trajectory.
This is demonstrated in Appendix \ref{appendix_white}. (\ref{Dw_integral}) is just} the Wigner transform of $D(t,t')$ where $\omega=0$ such that $D^w$ is independent of $\omega$; hence the name ``white-noise'' diffusion. Thus, the white-noise diffusion coefficient contains information about the correlations in the stochastic force but applies that information in a Markovian manner. A time-scale separation can be applied to the white-noise diffusion in a similar way as previously, whereby the adiabatic white-noise diffusion is given by {  
\begin{equation}
D^w_{(0)}(t_f) = \int^{t_f}_{-t_f} d\tau D_{(0)}(t_f,\tau ).
\end{equation}
}
The adiabatic white-noise diffusion is the most commonly used approximation to the diffusion coefficient\cite{preston2020,preston21,Bode12} and we note in passing that it satisfies a fluctuation-dissipation relation with (\ref{singleLevelVisc}) when $\partial_\nu h = \lambda$.
The white-noise approximation is valid when correlations in the exact diffusion coefficient decay over time-scales in which the effects of other forces present in the system (in our case, the Ehrenfest and the classical forces) are negligible. In order to quantify its validity, we introduce the correlation time defined according to {  
\begin{equation}
t_{\text{corr}}(t_f) = \frac{\int^{t_f}_{-t_f} d\tau D(t_f,\tau)}{D(t_f,\tau=0)}.
\label{tcorr}
\end{equation}}
This is a measure of the persistence of correlations in the stochastic force, independent of their strength. {  The numerator of (\ref{tcorr}) is our expression for the white-noise diffusion while the denominator is the variance in the stochastic force at time $t_f$.}
With all quantities now defined, we can begin to discuss the results.

\begin{figure*}
\begin{subfigure}{0.48\textwidth}
\centering
\includegraphics[width=1\textwidth]{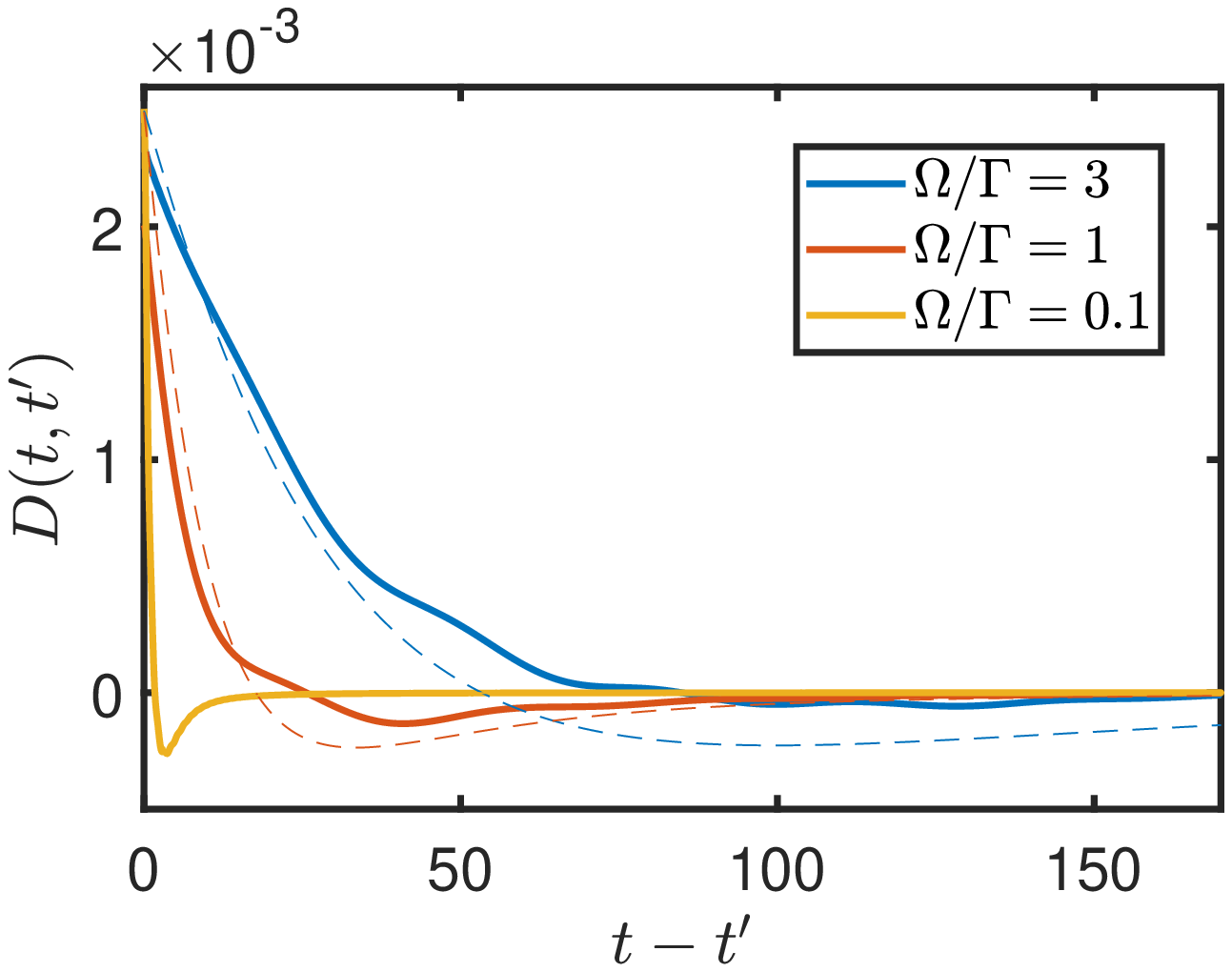}
\caption{}
\label{corr_a}
\end{subfigure}
\begin{subfigure}{0.48\textwidth}
\centering
\includegraphics[width=1\textwidth]{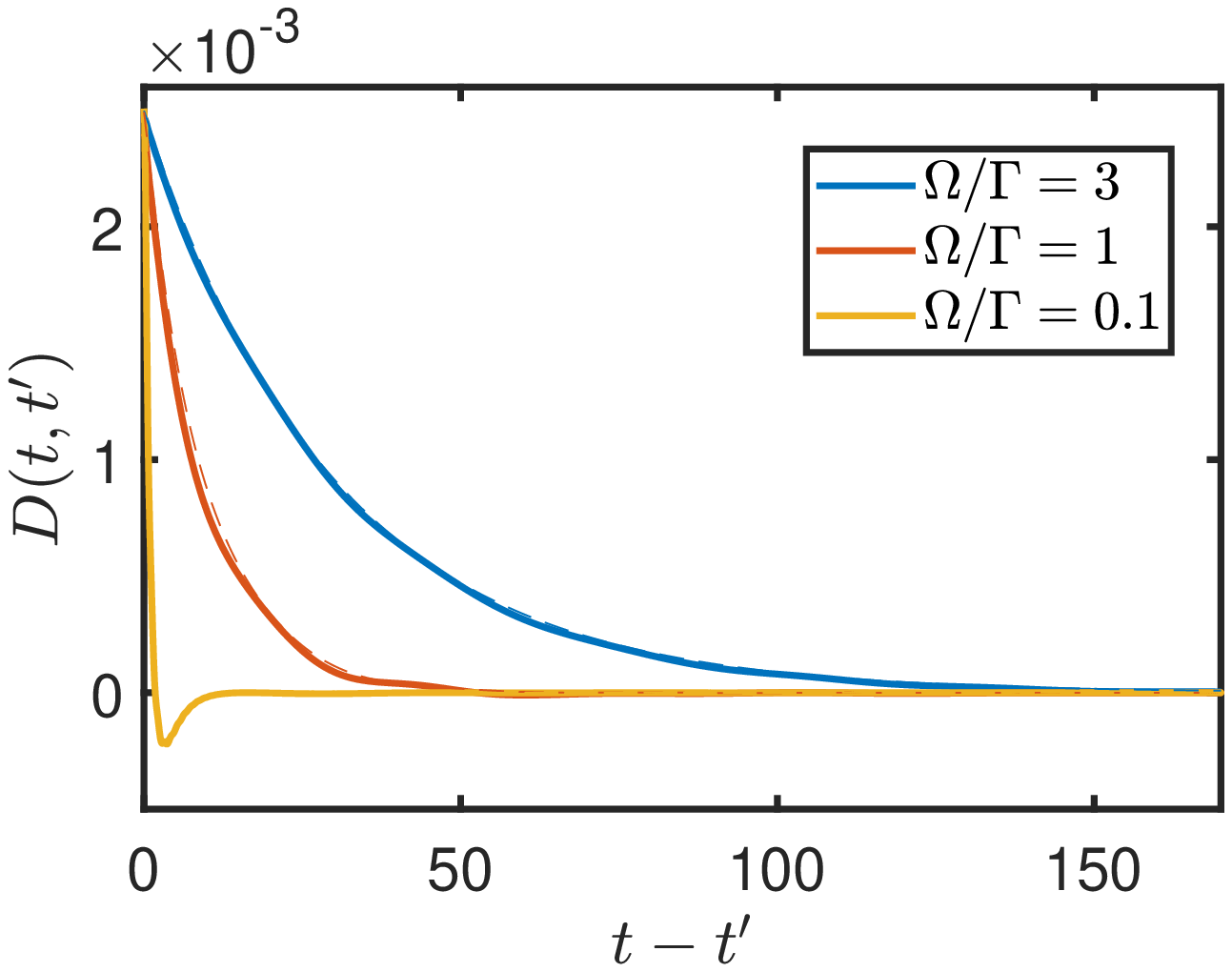}
\caption{}
\label{corr_b}
\end{subfigure}
\begin{subfigure}{0.48\textwidth}
\centering
\includegraphics[width=1\textwidth]{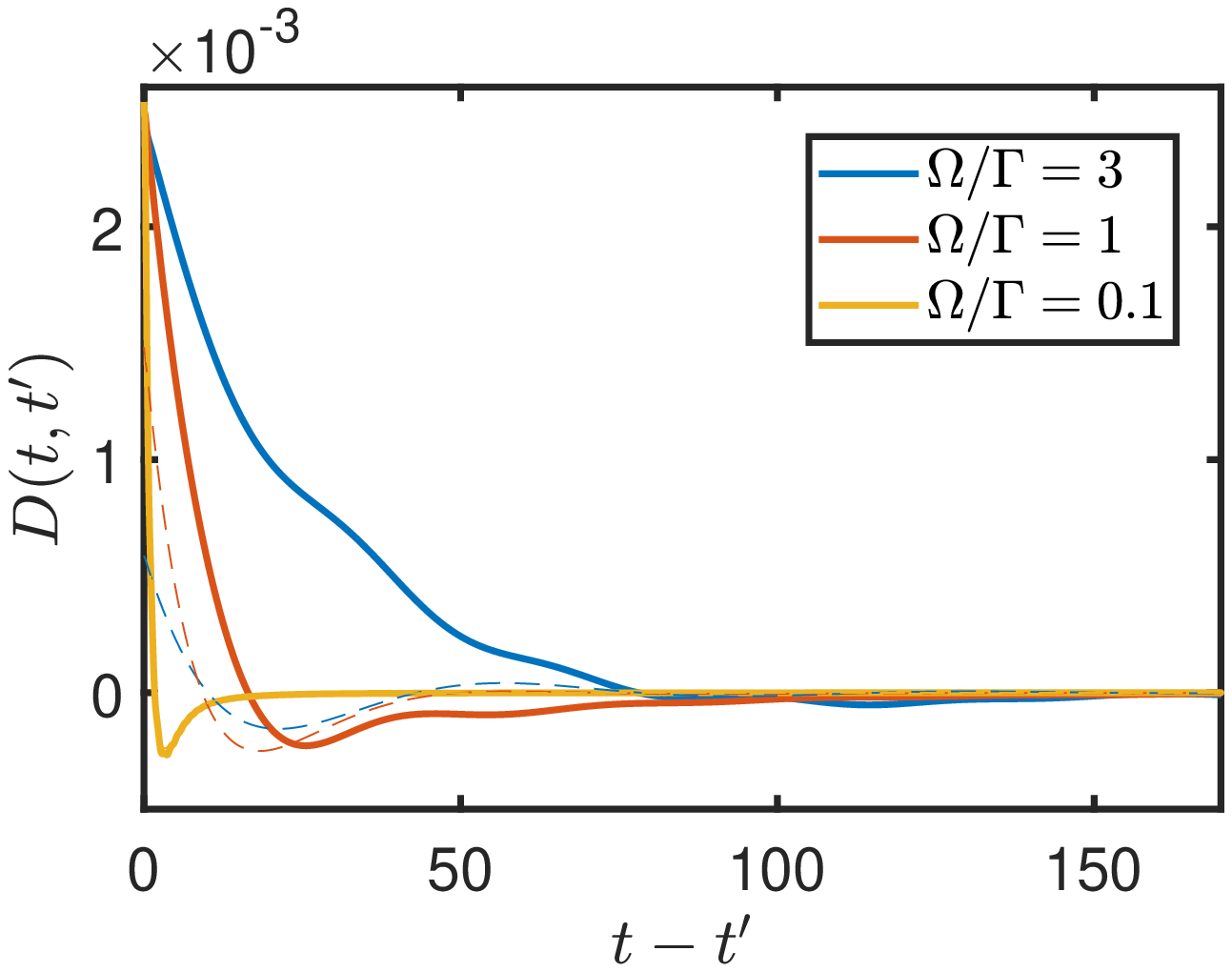}
\caption{}
\label{corr_c}
\end{subfigure}
\begin{subfigure}{0.48\textwidth}
\centering
\includegraphics[width=1\textwidth]{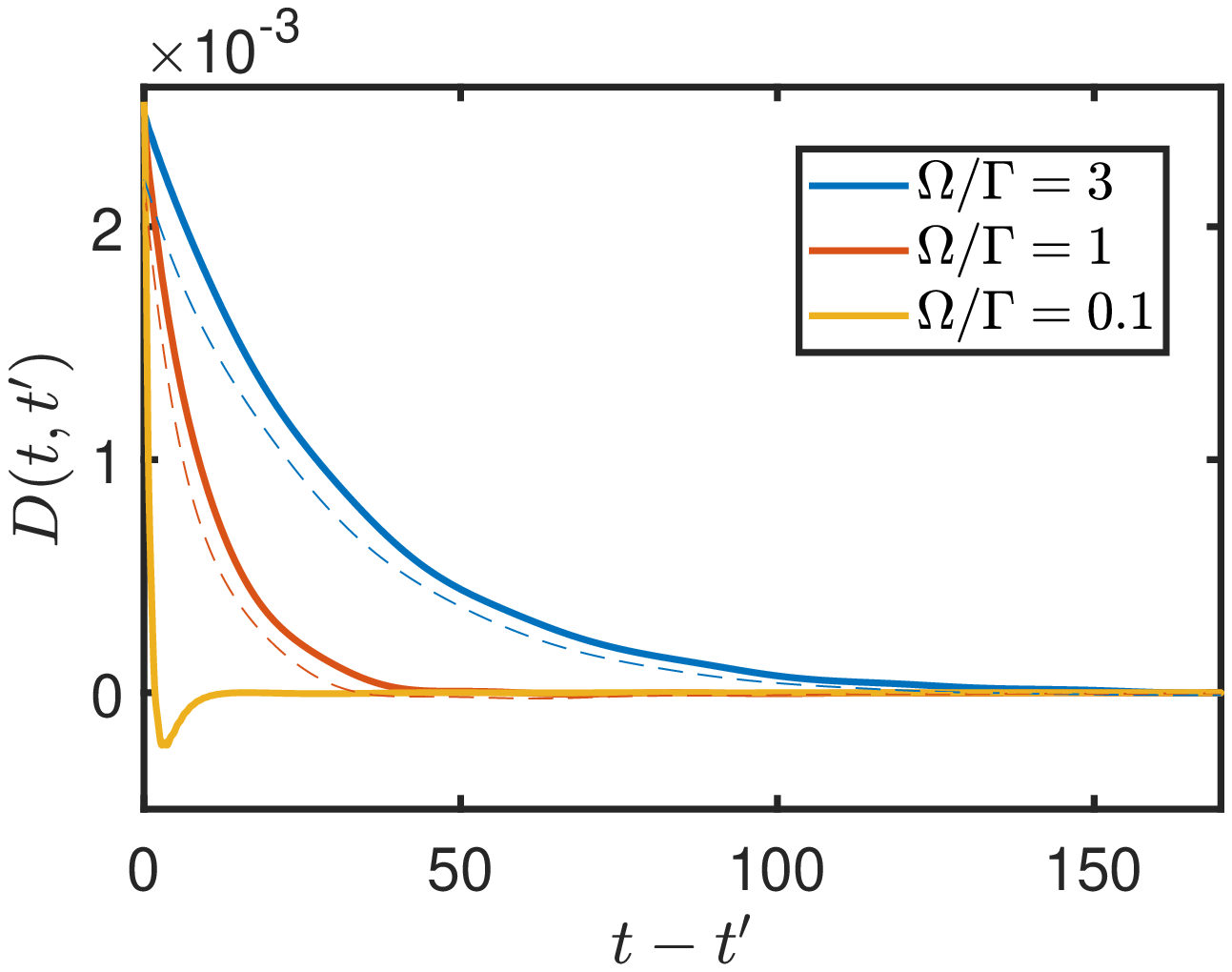}
\caption{}
\label{corr_d}
\end{subfigure}
\caption{Exact diffusion coefficient as a function of $\tau$ for where time $t$ occurs at: (a)/(b) $x=0$, (c)/(d) a turning point. Left: $V=0$, Right: $V = 0.2$. Dashed line is the corresponding $D_{(0)}(t,t')$ for the same parameters.}
\label{Dplots}
\end{figure*}

We once again apply the time-stepping algorithm presented in Section \ref{algorithm} to find classical Ehrenfest trajectories in time for our classical coordinate. In doing so, we store $A(\omega,t)$  (\ref{A}) at each time-step. The two-time lesser and greater Green's functions can then be computed according to (\ref{Glesser}) and (\ref{Ggreater}) by {  inputting} $A(\omega,t)$ at different points in the stored trajectory and numerically integrating over $\omega$ for each possible value of $\tau$. We then calculate the exact diffusion coefficient as a function of $\tau$ according to (\ref{our_D}), where $D(\tau = 0)$ corresponds to the variance in the stochastic force at the end point of the trajectory and $D(\tau>0)$ is the correlation in the stochastic force between the times $t$ and $t-\tau$. {  To clarify the method, the stochastic force is not included in the simulation. We instead calculate the trajectory using Ehrenfest dynamics as a means to assess the behaviour of the exact diffusion coefficient.} The common parameters in this section are $\lambda = 0.1$, $k = 1$ and $x_0 = 0$.

In Figure \ref{Dplots} (a)-(d) we observe the exact diffusion as in (\ref{our_D}) for different values of our small parameter (solid line) plotted against the corresponding adiabatic diffusion as per (\ref{D_ad}) (dashed line). Here, time $t$ corresponds to the end point of the trajectory where in (a)/(b) the trajectory is ended at a point when $x=0$, while in (c)/(d) the trajectory is ended at a turning point. The left plots are calculated in equilibrium while the right plots are calculated at a non-zero voltage. The differences between the solid and dashed lines give an indication of the effects of the feedback on the electronic environment due to the motion of the classical coordinate. {  To our knowledge, the effects of non-adiabatic motion on the diffusion coefficient have not been directly observed previously.} As expected, this feedback becomes especially important when the small parameter $\Omega/\Gamma$ becomes larger, such that the perturbative truncation of (\ref{our_D}) is no longer satisfied. However, we observe that an increase to the voltage nullifies the effects of the feedback. This is justified by the knowledge that the electronic tunnelling time-scale is faster at higher voltages\cite{cuevasbook}, meaning that the electronic environment can more readily equilibrate to any changes in the classical geometry. Thus, the time-scale separation becomes increasingly justifiable further from equilibrium, even despite $\Omega / \Gamma$ being large. We also find that the correlations in the stochastic force are dissipated over shorter time-scales when $\Omega / \Gamma$ is smaller such that $D(\tau)$ approaches a shape more reminiscent of a delta function. Finally, we note that the time-scale separation appears less satisfactory at the turning points of the trajectory. The acceleration of the classical coordinate is largest around the turning points and we posit that acceleration dependent terms become important here, which are otherwise unaccounted for under the assumption of adiabatic motion.

The effectiveness of the time-scale separation for different ending positions along the trajectory is quantified in Figure \ref{x_colour}, in which we have calculated $D^w / D^w_{(0)}$ for different end positions along a period of the trajectory when $\Omega/\Gamma = 1$. For these parameters, the time-scale separation is ineffective as $D^w$ is over twice as large as $D^w_{(0)}$ at a minimum. We observe the adiabatic assumption to be at its weakest in the vicinity of the turning points.

\begin{figure}
\centering
\includegraphics[width=0.48\textwidth]{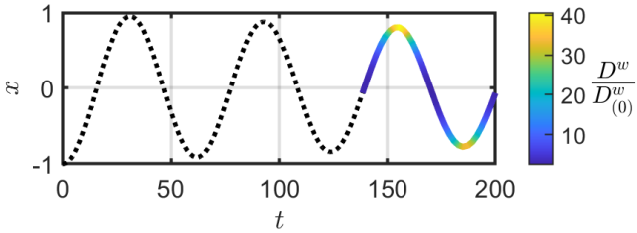}
\caption{An example trajectory of the classical coordinate. Colour scale shows the calculated $D^w / D^w_{(0)}$ at that point in the trajectory. Parameters: $\Omega / \Gamma = 1$, $V = 0$.}
\label{x_colour}
\end{figure}

In Figure \ref{corr_period}, we use $t_{\text{corr}}/T_p$ as a measure of the suitability of the white-noise approximation, where $T_p$ is the period of oscillations in the classical coordinate. Here, $t_\text{corr}$ contains any information about the electronic forces acting on the coordinate, while $T_p$ contains information on the classical force. We observe that an increase to $\Omega/\Gamma$ results in a corresponding increase to $t_\text{corr}/T_p$, implying that the white-noise approximation is more valid at smaller $\Omega/\Gamma$ where the time-scale separation is more well-defined. However, we note that the validity decreases upon increasing the voltage which would ordinarily serve to further increase the time-scale separation. {  For this system, a larger voltage results in more persistent correlations in the stochastic force which emerges in the numerator of (\ref{tcorr}), while the change in the denominator is comparatively irrelevant.}

\begin{figure}
\centering
\includegraphics[width=0.5\textwidth]{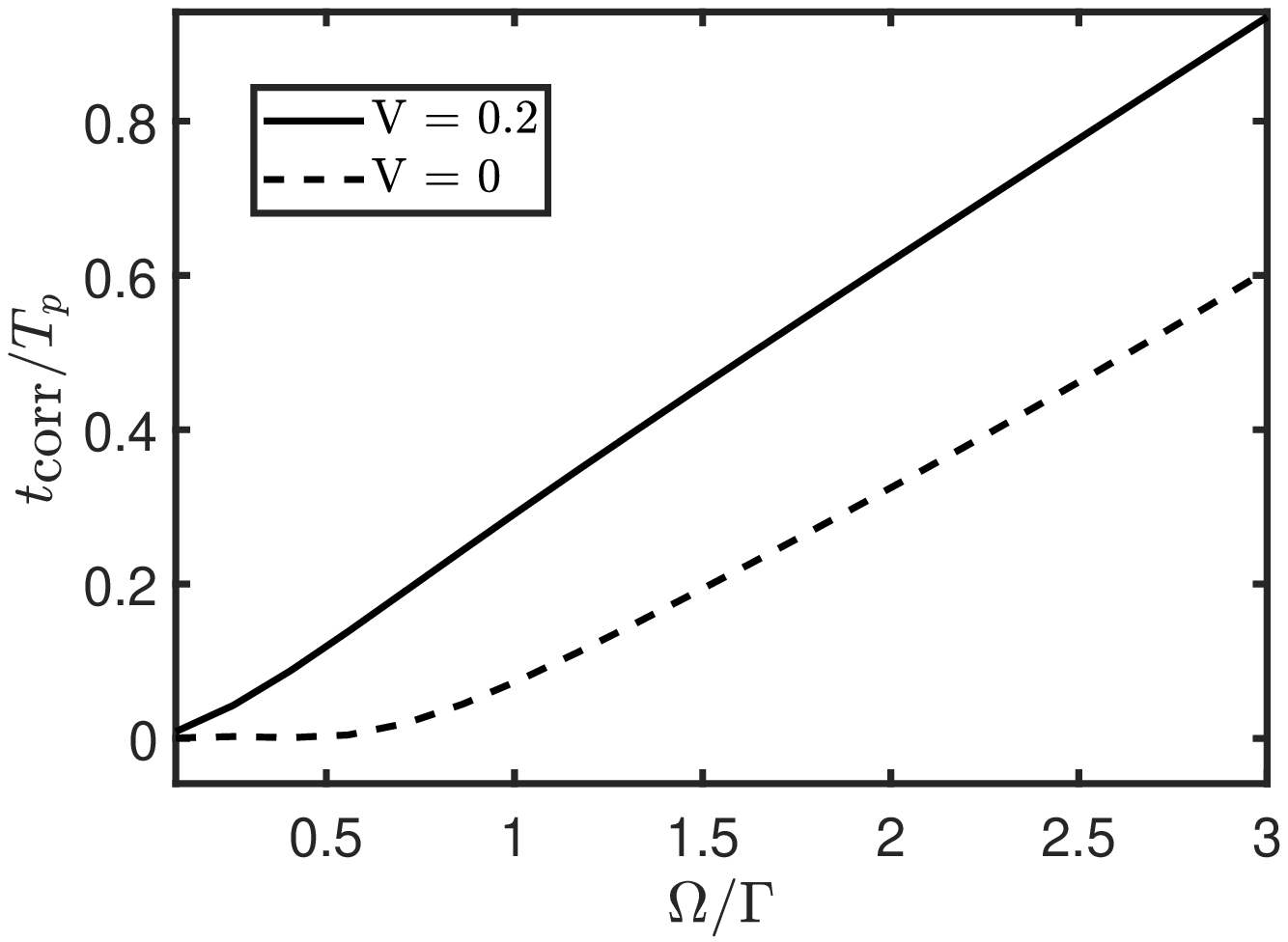}
\caption{The ratio of the correlation time to the period of classical oscillation, as a function of $\Omega / \Gamma$ for equilibrium and non-equilibrium cases. Trajectory length = 300.}
\label{corr_period}
\end{figure}

This counter-intuitive result also indirectly emerges in Figure \ref{corr_h0} where we observe the white-noise approximation to become more applicable outside of the voltage window and away from the chemical potentials of the leads. In this region, electrons tunnel more slowly through the central region. We do not yet have a convincing explanation for this.

\begin{figure}
\centering
\includegraphics[width=0.5\textwidth]{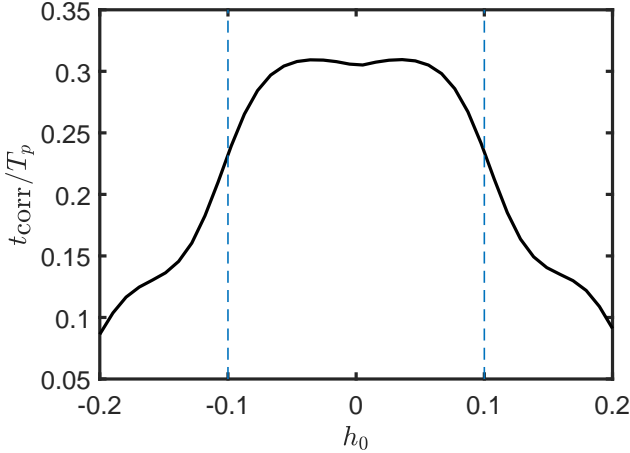}
\caption{The ratio of correlation time to the period of classical oscillation for shifted values of $h_0$. The dashed blue lines show the edges of the voltage window. Parameters: $\Omega / \Gamma = 1$.}
\label{corr_h0}
\end{figure}
}

\section{CONCLUSIONS}

In this paper, we have introduced a novel time-stepping algorithm for evaluating the exact lesser Green's function at equal times, which allows us to simulate the trajectory of multiple classical coordinates simultaneously via an Ehrenfest approach. We use this to benchmark the commonly used Langevin approach which necessitates the use of a time-scale separation between classical and electronic coordinates within the system. We observe that despite our avoidance of a time-scale separation within our system, we observe negative dissipations (positive power input to the classical coordinate) as predicted by the Langevin approach. We also note that these negative dissipations can be overwhelmed by positive dissipations due to higher order forces unaccounted for in the Langevin approach. {  We also apply our method to the calculation of the diffusion coefficient in which we observe the effect of the non-adiabatic feedback of the classical coordinate onto the electronic environment and its behaviour under a time-scale separation. Additionally, we assess the validity of the white-noise approximation for the diffusion coefficient for a range of parameters and find that it is most applicable under a clear time-scale separation within the system and is better applied when the transport channel energy is shifted outside of the voltage window.}

\appendix

\section{Viscosity Coefficient in the Single-Level Case}
\label{appendix_xi}

For the case of a single electronic level and any number of classical degrees of freedom, the first order force for an arbitrary degree of freedom $\nu$ in (\ref{order_forces}) is given by
\begin{equation}
F_{\nu , (1)} = \int \frac{d\omega}{2\pi} i\partial_\nu h \widetilde{G}_{(1)}^<,
\label{F1_single}
\end{equation}
where $\widetilde{G}_{(1)}^<$ in the single-level case can be simplified to 
\begin{equation}
\widetilde{G}_{(1)}^< = \frac{i}{2} \partial_T h \widetilde{G}_{(0)}^R \widetilde{G}_{(0)}^A \Big( \widetilde{G}_{(0)}^A - \widetilde{G}_{(0)}^R \Big) \partial_\omega \widetilde{\Sigma}^<.
\label{G1_single}
\end{equation}
Through the use of (\ref{GRA}), (\ref{sigma_lesser}), and the fact that $\partial_\omega f_\alpha = -\frac{1}{T_\alpha} f_\alpha(1-f_\alpha)$, some simplification yields
\begin{equation}
\widetilde{G}_{(1)}^< =  \frac{i\partial_T h \Gamma [\frac{f_L}{T_L}(1-f_L) \Gamma_L + \frac{f_R}{T_R} (1-f_R) \Gamma_R ]}{2\Big[ (\omega - h)^2 + \frac{\Gamma^2}{4} \Big]^2}.
\label{G1_simp}
\end{equation}
We obtain the simplified form of $F_{\nu , (1)}$ by substituting (\ref{G1_simp}) into (\ref{F1_single}):
\begin{equation}
F_{\nu , (1)} = - \partial_\nu h \partial_T h \Gamma  \int \frac{d\omega}{4\pi} \frac{\frac{f_L}{T_L} (1-f_L) \Gamma_L + \frac{f_R}{T_R} (1-f_R) \Gamma_R }{\Big[ (\omega - h)^2 + \frac{\Gamma^2}{4} \Big]^2}.
\end{equation}
We apply the chain rule such that $\partial_T h = \sum_{\nu '} v_{\nu '} \partial_{\nu '} h$ where $v_{\nu '}$ is the velocity of the $\nu '$ coordinate and the summation is over all classical degrees of freedom. The diagonal components of the force then correspond to when $\nu ' = \nu$, for which we will utilise an additional subscript:
\begin{equation}
F_{\nu \nu , (1)} = - v_\nu (\partial_\nu h)^2 \Gamma  \int \frac{d\omega}{4\pi} \frac{\frac{f_L}{T_L} (1-f_L) \Gamma_L + \frac{f_R}{T_R} (1-f_R) \Gamma_R }{\Big[ (\omega - h)^2 + \frac{\Gamma^2}{4} \Big]^2}.
\end{equation}
Finally, the diagonal components of the viscosity coefficient can then be found according to $\xi_{\nu \nu} = - \frac{F_{\nu \nu , (1)}}{v_\nu}$, which yields
\begin{equation}
\xi_{\nu \nu} = \frac{\Gamma(\partial_{\nu}h)^2}{4\pi}
\int d\omega \frac{\frac{f_L}{T_L}(1-f_L)\Gamma_L 
+ \frac{f_R}{T_R}(1-f_R)\Gamma_R}{\Big[(\omega-h)^2+\Gamma^2/4\Big]^2}.
\end{equation}

\section{Exact Diffusion Coefficient}
\label{appendix_D}

Here we derive (\ref{our_D}) for the case of a single electronic level and single classical degree of freedom $x$. We express the quantum force acting on the classical coordinate due to the electronic environment according to a mean component which contains the deterministic forces, and a stochastic component which captures the probabilistic fluctuations about the mean:
\begin{equation}
\hat{f}(t) = f(t) + \delta \hat{f} (t).
\end{equation}
By multiplying this equation by itself at a different time $t'$ and utilising the fact that $\langle \delta \hat{f}(t) \rangle = 0$, we find
\begin{equation}
\langle \delta \hat{f}(t) \delta \hat{f}(t') \rangle = \langle \hat{f}(t) \hat{f}(t') \rangle - f(t)f(t'),
\label{force_corrs}
\end{equation}
where the quantum force operator according to (\ref{heisenberg}) is given by
\begin{equation}
\hat{f}(t) = -\partial_x U -\partial_x h d^\dag d.
\end{equation}
After inputting this expression into (\ref{force_corrs}) and applying Wick's theorem to decompose the strings of creation and annihilation operators, we observe that the terms in $f(t)f(t')$ cancel exactly with part of the decomposition of $\langle \hat{f}(t) \hat{f}(t') \rangle$. We are then left with
\begin{align}
\langle \delta \hat{f}(t) \delta \hat{f}(t') \rangle &= (\partial_x h)^2 \langle d^\dag (t) d(t') \rangle \langle d(t) d^\dag (t')  \rangle,
\\
&= \lambda^2 G^> (t,t') G^< (t',t).
\end{align}
Here we have used the fact that $\partial_x h = \lambda$ for our model and we have utilised the definitions for the lesser and greater Green's functions.

{  
\section{Wigner Transform of (\ref{our_D})}
\label{appendix_col}
Direct application of the Wigner transform to (\ref{our_D}) yields
\begin{equation}
\tilde{D}(\omega,T) = \lambda^2\int d\tau e^{i\omega\tau} G^>(t,t')G^<(t',t).
\end{equation}

Now, we replace $G^<$ in the time domain by the inverse Wigner transform of $\tilde{G}^<$ and simplify to find
\begin{align}
\tilde{D}(\omega,T) &= \frac{\lambda^2}{2\pi}\int d\tau e^{i\omega\tau} G^>(t,t')\int d\omega' e^{-i\omega'(-\tau)} \tilde{G}^<(\omega',T),
\\
&= \frac{\lambda^2}{2\pi}\int d\omega' \int d\tau e^{i(\omega+\omega')\tau} G^>(t,t') \tilde{G}^<(\omega',T),
\\
&= \frac{\lambda^2}{2\pi}\int d\omega' \tilde{G}^>(\omega + \omega',T) \tilde{G}^<(\omega',T),
\end{align}
where we have used the definition of the Wigner transform.

}
{  
\section{Equation for the white-noise diffusion}
\label{appendix_white}
In (\ref{Dw_orig_def}), we introduced the white-noise diffusion $D^w$ as a means of representing the colored-noise diffusion in a Markovian manner. We are then tasked with finding an equation for $D^w$ whose application will accurately reproduce the dynamics produced by the exact colored-noise diffusion. To do so, we consider a pedagogical example in which we neglect all forces except the stochastic force. The governing equation of motion is given by
\begin{equation}
m\frac{dv}{dt}=\delta f(t).
\end{equation}
This is solved for the velocity at time $t$ according to 
\begin{equation}
v(t) = v(0) + \frac{1}{m}\int^t_0 dt' \delta f(t').
\end{equation}
The change in kinetic energy of the classical coordinate is then found by squaring the above and taking an average over the fluctuations such that we obtain
\begin{equation}
\Delta KE = \frac{1}{2m} \int^t_0 dt' \int^t_0 dt'' D(t',t'').
\label{KE_full}
\end{equation}
Now, we implement the white noise approximation according to (\ref{Dw_orig_def}) and solve for an equation for $D^w$ which yields the same change in kinetic energy. Making the transformation from $(t',t'')$ to $(T,\tau )$ yields
\begin{align}
\Delta KE &= \frac{1}{2m} \int^t_0 dT \int^t_{-t} d\tau D^w (T)\delta(\tau),\\
&= \frac{1}{2m} \int^t_0 dT  D^w (T).
\label{KE_white}
\end{align}
By enforcing that (\ref{KE_white}) and (\ref{KE_full}) are equal, we find
\begin{equation}
\int^t_0 dT  D^w (T) = \int^t_0 dT \int^t_{-t} d\tau D(T,\tau).
\end{equation}
Application of $\frac{d}{dt}$ to both sides and using the symmetry property, $D(T,\tau)=D(T,-\tau)$, then yields
\begin{equation}
D^w (t) = \int^t_{-t} d\tau D(t,\tau) + 2\int^t_0 dT D(T,t).
\end{equation}
Finally, we assume that time $t$ is sufficiently large such that $D(T,t)$ is approximately zero. This amounts to assuming that the correlations in the stochastic force decay to zero over time intervals of $t$ or longer. The second term disappears and we are left with our final expression for the white-noise diffusion coefficient:
\begin{equation}
D^w (t) = \int^t_{-t} d\tau D(t,\tau).
\end{equation}
Clearly, the white noise approximation in this case will produce the same observable change in kinetic energy over any time-scale. However, the inclusion of a frictional force and an external potential will limit the validity of the approximation to finite time-scales dependent on their respective strengths. This is because these forces may produce an appreciable effect on the dynamics over the time-scales for which the stochastic force is correlated.

}

\clearpage
\bibliography{paperbib.bib,additional.bib}

\begin{thebibliography}{76}%
\makeatletter
\providecommand \@ifxundefined [1]{%
 \@ifx{#1\undefined}
}%
\providecommand \@ifnum [1]{%
 \ifnum #1\expandafter \@firstoftwo
 \else \expandafter \@secondoftwo
 \fi
}%
\providecommand \@ifx [1]{%
 \ifx #1\expandafter \@firstoftwo
 \else \expandafter \@secondoftwo
 \fi
}%
\providecommand \natexlab [1]{#1}%
\providecommand \enquote  [1]{``#1''}%
\providecommand \bibnamefont  [1]{#1}%
\providecommand \bibfnamefont [1]{#1}%
\providecommand \citenamefont [1]{#1}%
\providecommand \href@noop [0]{\@secondoftwo}%
\providecommand \href [0]{\begingroup \@sanitize@url \@href}%
\providecommand \@href[1]{\@@startlink{#1}\@@href}%
\providecommand \@@href[1]{\endgroup#1\@@endlink}%
\providecommand \@sanitize@url [0]{\catcode `\\12\catcode `\$12\catcode
  `\&12\catcode `\#12\catcode `\^12\catcode `\_12\catcode `\%12\relax}%
\providecommand \@@startlink[1]{}%
\providecommand \@@endlink[0]{}%
\providecommand \url  [0]{\begingroup\@sanitize@url \@url }%
\providecommand \@url [1]{\endgroup\@href {#1}{\urlprefix }}%
\providecommand \urlprefix  [0]{URL }%
\providecommand \Eprint [0]{\href }%
\providecommand \doibase [0]{https://doi.org/}%
\providecommand \selectlanguage [0]{\@gobble}%
\providecommand \bibinfo  [0]{\@secondoftwo}%
\providecommand \bibfield  [0]{\@secondoftwo}%
\providecommand \translation [1]{[#1]}%
\providecommand \BibitemOpen [0]{}%
\providecommand \bibitemStop [0]{}%
\providecommand \bibitemNoStop [0]{.\EOS\space}%
\providecommand \EOS [0]{\spacefactor3000\relax}%
\providecommand \BibitemShut  [1]{\csname bibitem#1\endcsname}%
\let\auto@bib@innerbib\@empty
\bibitem [{\citenamefont {Cuevas}\ and\ \citenamefont
  {Scheer}(2017)}]{cuevasbook}%
  \BibitemOpen
  \bibfield  {author} {\bibinfo {author} {\bibfnamefont {J.~C.}\ \bibnamefont
  {Cuevas}}\ and\ \bibinfo {author} {\bibfnamefont {E.}~\bibnamefont
  {Scheer}},\ }\href {https://doi.org/10.1142/10598} {\emph {\bibinfo {title}
  {{Molecular Electronics}}}},\ \bibinfo {series} {World Scientific Series in
  Nanoscience and Nanotechnology}, Vol.~\bibinfo {volume} {15}\ (\bibinfo
  {publisher} {WORLD SCIENTIFIC},\ \bibinfo {year} {2017})\BibitemShut
  {NoStop}%
\bibitem [{\citenamefont {Ward}\ \emph {et~al.}(2011)\citenamefont {Ward},
  \citenamefont {Corley}, \citenamefont {Tour},\ and\ \citenamefont
  {Natelson}}]{heating11}%
  \BibitemOpen
  \bibfield  {author} {\bibinfo {author} {\bibfnamefont {D.~R.}\ \bibnamefont
  {Ward}}, \bibinfo {author} {\bibfnamefont {D.~A.}\ \bibnamefont {Corley}},
  \bibinfo {author} {\bibfnamefont {J.~M.}\ \bibnamefont {Tour}},\ and\
  \bibinfo {author} {\bibfnamefont {D.}~\bibnamefont {Natelson}},\ }\bibfield
  {title} {\bibinfo {title} {{Vibrational and electronic heating in nanoscale
  junctions}},\ }\href@noop {} {\bibfield  {journal} {\bibinfo  {journal}
  {Nature Nanotechnology}\ }\textbf {\bibinfo {volume} {6}} (\bibinfo {year}
  {2011})}\BibitemShut {NoStop}%
\bibitem [{\citenamefont {Sabater}\ \emph {et~al.}(2015)\citenamefont
  {Sabater}, \citenamefont {Untiedt},\ and\ \citenamefont {van
  Ruitenbeek}}]{Sabater15}%
  \BibitemOpen
  \bibfield  {author} {\bibinfo {author} {\bibfnamefont {C.}~\bibnamefont
  {Sabater}}, \bibinfo {author} {\bibfnamefont {C.}~\bibnamefont {Untiedt}},\
  and\ \bibinfo {author} {\bibfnamefont {J.~M.}\ \bibnamefont {van
  Ruitenbeek}},\ }\bibfield  {title} {\bibinfo {title} {{Evidence for
  non-conservative current-induced forces in the breaking of Au and Pt atomic
  chains}},\ }\href@noop {} {\bibfield  {journal} {\bibinfo  {journal}
  {Beilstein J. Nanotechnol.}\ }\textbf {\bibinfo {volume} {6}},\ \bibinfo
  {pages} {2338} (\bibinfo {year} {2015})}\BibitemShut {NoStop}%
\bibitem [{\citenamefont {Tsutsui}\ \emph {et~al.}(2008)\citenamefont
  {Tsutsui}, \citenamefont {Taniguchi},\ and\ \citenamefont
  {Kawai}}]{Tsutsui08}%
  \BibitemOpen
  \bibfield  {author} {\bibinfo {author} {\bibfnamefont {M.}~\bibnamefont
  {Tsutsui}}, \bibinfo {author} {\bibfnamefont {M.}~\bibnamefont {Taniguchi}},\
  and\ \bibinfo {author} {\bibfnamefont {T.}~\bibnamefont {Kawai}},\ }\bibfield
   {title} {\bibinfo {title} {{Local Heating in Metal-Molecule-Metal
  Junctions}},\ }\href {https://doi.org/10.1021/nl801669e} {\bibfield
  {journal} {\bibinfo  {journal} {Nano Letters}\ }\textbf {\bibinfo {volume}
  {8}},\ \bibinfo {pages} {3293} (\bibinfo {year} {2008})}\BibitemShut
  {NoStop}%
\bibitem [{\citenamefont {Schulze}\ \emph
  {et~al.}(2008{\natexlab{a}})\citenamefont {Schulze}, \citenamefont {Franke},
  \citenamefont {Gagliardi}, \citenamefont {Romano}, \citenamefont {Lin},
  \citenamefont {Rosa}, \citenamefont {Niehaus}, \citenamefont {Frauenheim},
  \citenamefont {{Di Carlo}}, \citenamefont {Pecchia},\ and\ \citenamefont
  {Pascual}}]{Schulze08}%
  \BibitemOpen
  \bibfield  {author} {\bibinfo {author} {\bibfnamefont {G.}~\bibnamefont
  {Schulze}}, \bibinfo {author} {\bibfnamefont {K.~J.}\ \bibnamefont {Franke}},
  \bibinfo {author} {\bibfnamefont {A.}~\bibnamefont {Gagliardi}}, \bibinfo
  {author} {\bibfnamefont {G.}~\bibnamefont {Romano}}, \bibinfo {author}
  {\bibfnamefont {C.~S.}\ \bibnamefont {Lin}}, \bibinfo {author} {\bibfnamefont
  {A.~L.}\ \bibnamefont {Rosa}}, \bibinfo {author} {\bibfnamefont {T.~A.}\
  \bibnamefont {Niehaus}}, \bibinfo {author} {\bibfnamefont {T.}~\bibnamefont
  {Frauenheim}}, \bibinfo {author} {\bibfnamefont {A.}~\bibnamefont {{Di
  Carlo}}}, \bibinfo {author} {\bibfnamefont {A.}~\bibnamefont {Pecchia}},\
  and\ \bibinfo {author} {\bibfnamefont {J.~I.}\ \bibnamefont {Pascual}},\
  }\bibfield  {title} {\bibinfo {title} {{Resonant Electron Heating and
  Molecular Phonon Cooling in Single ${\mathrm{C}}_{60}$ Junctions}},\ }\href
  {https://doi.org/10.1103/PhysRevLett.100.136801} {\bibfield  {journal}
  {\bibinfo  {journal} {Phys. Rev. Lett.}\ }\textbf {\bibinfo {volume} {100}},\
  \bibinfo {pages} {136801} (\bibinfo {year} {2008}{\natexlab{a}})}\BibitemShut
  {NoStop}%
\bibitem [{\citenamefont {Ioffe}\ \emph {et~al.}(2008)\citenamefont {Ioffe},
  \citenamefont {Shamai}, \citenamefont {Ophir}, \citenamefont {Noy},
  \citenamefont {Yutsis}, \citenamefont {Kfir}, \citenamefont {Cheshnovsky},\
  and\ \citenamefont {Selzer}}]{Ioffe08}%
  \BibitemOpen
  \bibfield  {author} {\bibinfo {author} {\bibfnamefont {Z.}~\bibnamefont
  {Ioffe}}, \bibinfo {author} {\bibfnamefont {T.}~\bibnamefont {Shamai}},
  \bibinfo {author} {\bibfnamefont {A.}~\bibnamefont {Ophir}}, \bibinfo
  {author} {\bibfnamefont {G.}~\bibnamefont {Noy}}, \bibinfo {author}
  {\bibfnamefont {I.}~\bibnamefont {Yutsis}}, \bibinfo {author} {\bibfnamefont
  {K.}~\bibnamefont {Kfir}}, \bibinfo {author} {\bibfnamefont {O.}~\bibnamefont
  {Cheshnovsky}},\ and\ \bibinfo {author} {\bibfnamefont {Y.}~\bibnamefont
  {Selzer}},\ }\bibfield  {title} {\bibinfo {title} {{Detection of heating in
  current-carrying molecular junctions by Raman scattering}},\ }\href@noop {}
  {\bibfield  {journal} {\bibinfo  {journal} {Nat Nano}\ }\textbf {\bibinfo
  {volume} {3}},\ \bibinfo {pages} {727} (\bibinfo {year} {2008})}\BibitemShut
  {NoStop}%
\bibitem [{\citenamefont {Franke}\ and\ \citenamefont
  {Pascual}(2012)}]{Franke12}%
  \BibitemOpen
  \bibfield  {author} {\bibinfo {author} {\bibfnamefont {K.~J.}\ \bibnamefont
  {Franke}}\ and\ \bibinfo {author} {\bibfnamefont {J.~I.}\ \bibnamefont
  {Pascual}},\ }\bibfield  {title} {\bibinfo {title} {{Effects of
  electron–vibration coupling in transport through single molecules}},\
  }\href {https://doi.org/10.1088/0953-8984/24/39/394002} {\bibfield  {journal}
  {\bibinfo  {journal} {Journal of Physics: Condensed Matter}\ }\textbf
  {\bibinfo {volume} {24}},\ \bibinfo {pages} {394002} (\bibinfo {year}
  {2012})}\BibitemShut {NoStop}%
\bibitem [{\citenamefont {Mukherjee}\ \emph {et~al.}(2013)\citenamefont
  {Mukherjee}, \citenamefont {Libisch}, \citenamefont {Large}, \citenamefont
  {Neumann}, \citenamefont {Brown}, \citenamefont {Cheng}, \citenamefont
  {Lassiter}, \citenamefont {Carter}, \citenamefont {Nordlander},\ and\
  \citenamefont {Halas}}]{Mukherjee13}%
  \BibitemOpen
  \bibfield  {author} {\bibinfo {author} {\bibfnamefont {S.}~\bibnamefont
  {Mukherjee}}, \bibinfo {author} {\bibfnamefont {F.}~\bibnamefont {Libisch}},
  \bibinfo {author} {\bibfnamefont {N.}~\bibnamefont {Large}}, \bibinfo
  {author} {\bibfnamefont {O.}~\bibnamefont {Neumann}}, \bibinfo {author}
  {\bibfnamefont {L.~V.}\ \bibnamefont {Brown}}, \bibinfo {author}
  {\bibfnamefont {J.}~\bibnamefont {Cheng}}, \bibinfo {author} {\bibfnamefont
  {J.~B.}\ \bibnamefont {Lassiter}}, \bibinfo {author} {\bibfnamefont {E.~A.}\
  \bibnamefont {Carter}}, \bibinfo {author} {\bibfnamefont {P.}~\bibnamefont
  {Nordlander}},\ and\ \bibinfo {author} {\bibfnamefont {N.~J.}\ \bibnamefont
  {Halas}},\ }\bibfield  {title} {\bibinfo {title} {{Hot Electrons Do the
  Impossible: Plasmon-Induced Dissociation of H2 on Au}},\ }\href
  {https://doi.org/10.1021/nl303940z} {\bibfield  {journal} {\bibinfo
  {journal} {Nano Letters}\ }\textbf {\bibinfo {volume} {13}},\ \bibinfo
  {pages} {240} (\bibinfo {year} {2013})}\BibitemShut {NoStop}%
\bibitem [{\citenamefont {Huang}\ \emph {et~al.}(2006)\citenamefont {Huang},
  \citenamefont {Xu}, \citenamefont {Chen}, \citenamefont {Ventra},\ and\
  \citenamefont {Tao}}]{Huang06}%
  \BibitemOpen
  \bibfield  {author} {\bibinfo {author} {\bibfnamefont {Z.}~\bibnamefont
  {Huang}}, \bibinfo {author} {\bibfnamefont {B.}~\bibnamefont {Xu}}, \bibinfo
  {author} {\bibfnamefont {Y.}~\bibnamefont {Chen}}, \bibinfo {author}
  {\bibfnamefont {M.~D.}\ \bibnamefont {Ventra}},\ and\ \bibinfo {author}
  {\bibfnamefont {N.~J.}\ \bibnamefont {Tao}},\ }\bibfield  {title} {\bibinfo
  {title} {{Measurement of Current-Induced Local Heating in a Single Molecule
  Junction}},\ }\href {https://doi.org/10.1021/nl0608285} {\bibfield  {journal}
  {\bibinfo  {journal} {Nano Letters}\ }\textbf {\bibinfo {volume} {6}},\
  \bibinfo {pages} {1240} (\bibinfo {year} {2006})}\BibitemShut {NoStop}%
\bibitem [{\citenamefont {de~Leon}\ \emph {et~al.}(2008)\citenamefont
  {de~Leon}, \citenamefont {Liang}, \citenamefont {Gu},\ and\ \citenamefont
  {Park}}]{deLeon08}%
  \BibitemOpen
  \bibfield  {author} {\bibinfo {author} {\bibfnamefont {N.~P.}\ \bibnamefont
  {de~Leon}}, \bibinfo {author} {\bibfnamefont {W.}~\bibnamefont {Liang}},
  \bibinfo {author} {\bibfnamefont {Q.}~\bibnamefont {Gu}},\ and\ \bibinfo
  {author} {\bibfnamefont {H.}~\bibnamefont {Park}},\ }\bibfield  {title}
  {\bibinfo {title} {{Vibrational Excitation in Single-Molecule Transistors:
  Deviation from the Simple Franck-Condon Prediction}},\ }\href
  {https://doi.org/10.1021/nl8018824} {\bibfield  {journal} {\bibinfo
  {journal} {Nano Letters}\ }\textbf {\bibinfo {volume} {8}},\ \bibinfo {pages}
  {2963} (\bibinfo {year} {2008})}\BibitemShut {NoStop}%
\bibitem [{\citenamefont {H{\"{u}}ttel}\ \emph {et~al.}(2009)\citenamefont
  {H{\"{u}}ttel}, \citenamefont {Witkamp}, \citenamefont {Leijnse},
  \citenamefont {Wegewijs},\ and\ \citenamefont {van~der Zant}}]{Huttel09}%
  \BibitemOpen
  \bibfield  {author} {\bibinfo {author} {\bibfnamefont {A.~K.}\ \bibnamefont
  {H{\"{u}}ttel}}, \bibinfo {author} {\bibfnamefont {B.}~\bibnamefont
  {Witkamp}}, \bibinfo {author} {\bibfnamefont {M.}~\bibnamefont {Leijnse}},
  \bibinfo {author} {\bibfnamefont {M.~R.}\ \bibnamefont {Wegewijs}},\ and\
  \bibinfo {author} {\bibfnamefont {H.~S.~J.}\ \bibnamefont {van~der Zant}},\
  }\bibfield  {title} {\bibinfo {title} {{Pumping of Vibrational Excitations in
  the Coulomb-Blockade Regime in a Suspended Carbon Nanotube}},\ }\href
  {https://doi.org/10.1103/PhysRevLett.102.225501} {\bibfield  {journal}
  {\bibinfo  {journal} {Physical Review Letters}\ }\textbf {\bibinfo {volume}
  {102}},\ \bibinfo {pages} {225501} (\bibinfo {year} {2009})}\BibitemShut
  {NoStop}%
\bibitem [{\citenamefont {Stettenheim}\ \emph {et~al.}(2010)\citenamefont
  {Stettenheim}, \citenamefont {Thalakulam}, \citenamefont {Pan},\ and\
  \citenamefont {Others}}]{Stettenheim10}%
  \BibitemOpen
  \bibfield  {author} {\bibinfo {author} {\bibfnamefont {J.}~\bibnamefont
  {Stettenheim}}, \bibinfo {author} {\bibfnamefont {M.}~\bibnamefont
  {Thalakulam}}, \bibinfo {author} {\bibfnamefont {F.}~\bibnamefont {Pan}},\
  and\ \bibinfo {author} {\bibnamefont {Others}},\ }\bibfield  {title}
  {\bibinfo {title} {{A macroscopic mechanical resonator driven by mesoscopic
  electrical back-action.}},\ }\href@noop {} {\bibfield  {journal} {\bibinfo
  {journal} {Nature}\ }\textbf {\bibinfo {volume} {466}},\ \bibinfo {pages}
  {86} (\bibinfo {year} {2010})}\BibitemShut {NoStop}%
\bibitem [{\citenamefont {Schulze}\ \emph
  {et~al.}(2008{\natexlab{b}})\citenamefont {Schulze}, \citenamefont {Franke},\
  and\ \citenamefont {Pascual}}]{Schulze08_2}%
  \BibitemOpen
  \bibfield  {author} {\bibinfo {author} {\bibfnamefont {G.}~\bibnamefont
  {Schulze}}, \bibinfo {author} {\bibfnamefont {K.~J.}\ \bibnamefont
  {Franke}},\ and\ \bibinfo {author} {\bibfnamefont {J.~I.}\ \bibnamefont
  {Pascual}},\ }\bibfield  {title} {\bibinfo {title} {{Resonant heating and
  substrate-mediated cooling of a single C 60 molecule in a tunnel junction}},\
  }\href {https://doi.org/10.1088/1367-2630/10/6/065005} {\bibfield  {journal}
  {\bibinfo  {journal} {New Journal of Physics}\ }\textbf {\bibinfo {volume}
  {10}},\ \bibinfo {pages} {065005} (\bibinfo {year}
  {2008}{\natexlab{b}})}\BibitemShut {NoStop}%
\bibitem [{\citenamefont {Hahn}\ and\ \citenamefont {Ho}(2005)}]{Hahn05}%
  \BibitemOpen
  \bibfield  {author} {\bibinfo {author} {\bibfnamefont {J.~R.}\ \bibnamefont
  {Hahn}}\ and\ \bibinfo {author} {\bibfnamefont {W.}~\bibnamefont {Ho}},\
  }\bibfield  {title} {\bibinfo {title} {{Chemisorption and dissociation of
  single oxygen molecules on Ag(110)}},\ }\href
  {https://doi.org/10.1063/1.2131064} {\bibfield  {journal} {\bibinfo
  {journal} {The Journal of Chemical Physics}\ }\textbf {\bibinfo {volume}
  {123}},\ \bibinfo {pages} {214702} (\bibinfo {year} {2005})}\BibitemShut
  {NoStop}%
\bibitem [{\citenamefont {Roy}\ \emph {et~al.}(2013)\citenamefont {Roy},
  \citenamefont {Mujica},\ and\ \citenamefont {Ratner}}]{ratner13}%
  \BibitemOpen
  \bibfield  {author} {\bibinfo {author} {\bibfnamefont {S.}~\bibnamefont
  {Roy}}, \bibinfo {author} {\bibfnamefont {V.}~\bibnamefont {Mujica}},\ and\
  \bibinfo {author} {\bibfnamefont {M.~A.}\ \bibnamefont {Ratner}},\ }\bibfield
   {title} {\bibinfo {title} {{Chemistry at molecular junctions: Rotation and
  dissociation of O2 on the Ag(110) surface induced by a scanning tunneling
  microscope}},\ }\href {https://doi.org/10.1063/1.4818163} {\bibfield
  {journal} {\bibinfo  {journal} {The Journal of Chemical Physics}\ }\textbf
  {\bibinfo {volume} {139}},\ \bibinfo {pages} {74702} (\bibinfo {year}
  {2013})}\BibitemShut {NoStop}%
\bibitem [{\citenamefont {Hla}\ \emph {et~al.}(2000)\citenamefont {Hla},
  \citenamefont {Bartels}, \citenamefont {Meyer},\ and\ \citenamefont
  {Rieder}}]{Hla00}%
  \BibitemOpen
  \bibfield  {author} {\bibinfo {author} {\bibfnamefont {S.-W.}\ \bibnamefont
  {Hla}}, \bibinfo {author} {\bibfnamefont {L.}~\bibnamefont {Bartels}},
  \bibinfo {author} {\bibfnamefont {G.}~\bibnamefont {Meyer}},\ and\ \bibinfo
  {author} {\bibfnamefont {K.-H.}\ \bibnamefont {Rieder}},\ }\bibfield  {title}
  {\bibinfo {title} {{Inducing All Steps of a Chemical Reaction with the
  Scanning Tunneling Microscope Tip: Towards Single Molecule Engineering}},\
  }\href {https://doi.org/10.1103/PhysRevLett.85.2777} {\bibfield  {journal}
  {\bibinfo  {journal} {Phys. Rev. Lett.}\ }\textbf {\bibinfo {volume} {85}},\
  \bibinfo {pages} {2777} (\bibinfo {year} {2000})}\BibitemShut {NoStop}%
\bibitem [{\citenamefont {Lauhon}\ and\ \citenamefont {Ho}(2000)}]{Lauhon00}%
  \BibitemOpen
  \bibfield  {author} {\bibinfo {author} {\bibfnamefont {L.~J.}\ \bibnamefont
  {Lauhon}}\ and\ \bibinfo {author} {\bibfnamefont {W.}~\bibnamefont {Ho}},\
  }\bibfield  {title} {\bibinfo {title} {{Control and Characterization of a
  Multistep Unimolecular Reaction}},\ }\href
  {https://doi.org/10.1103/PhysRevLett.84.1527} {\bibfield  {journal} {\bibinfo
   {journal} {Phys. Rev. Lett.}\ }\textbf {\bibinfo {volume} {84}},\ \bibinfo
  {pages} {1527} (\bibinfo {year} {2000})}\BibitemShut {NoStop}%
\bibitem [{\citenamefont {Stipe}\ \emph {et~al.}(1997)\citenamefont {Stipe},
  \citenamefont {Rezaei}, \citenamefont {Ho}, \citenamefont {Gao},
  \citenamefont {Persson},\ and\ \citenamefont {Lundqvist}}]{Stipe97}%
  \BibitemOpen
  \bibfield  {author} {\bibinfo {author} {\bibfnamefont {B.~C.}\ \bibnamefont
  {Stipe}}, \bibinfo {author} {\bibfnamefont {M.~A.}\ \bibnamefont {Rezaei}},
  \bibinfo {author} {\bibfnamefont {W.}~\bibnamefont {Ho}}, \bibinfo {author}
  {\bibfnamefont {S.}~\bibnamefont {Gao}}, \bibinfo {author} {\bibfnamefont
  {M.}~\bibnamefont {Persson}},\ and\ \bibinfo {author} {\bibfnamefont {B.~I.}\
  \bibnamefont {Lundqvist}},\ }\bibfield  {title} {\bibinfo {title}
  {{Single-Molecule Dissociation by Tunneling Electrons}},\ }\href
  {https://doi.org/10.1103/PhysRevLett.78.4410} {\bibfield  {journal} {\bibinfo
   {journal} {Phys. Rev. Lett.}\ }\textbf {\bibinfo {volume} {78}},\ \bibinfo
  {pages} {4410} (\bibinfo {year} {1997})}\BibitemShut {NoStop}%
\bibitem [{\citenamefont {Li}\ \emph {et~al.}(2015)\citenamefont {Li},
  \citenamefont {Su}, \citenamefont {Zhang}, \citenamefont {Steigerwald},
  \citenamefont {Nuckolls},\ and\ \citenamefont {Venkataraman}}]{Haixing15}%
  \BibitemOpen
  \bibfield  {author} {\bibinfo {author} {\bibfnamefont {H.}~\bibnamefont
  {Li}}, \bibinfo {author} {\bibfnamefont {T.~A.}\ \bibnamefont {Su}}, \bibinfo
  {author} {\bibfnamefont {V.}~\bibnamefont {Zhang}}, \bibinfo {author}
  {\bibfnamefont {M.~L.}\ \bibnamefont {Steigerwald}}, \bibinfo {author}
  {\bibfnamefont {C.}~\bibnamefont {Nuckolls}},\ and\ \bibinfo {author}
  {\bibfnamefont {L.}~\bibnamefont {Venkataraman}},\ }\bibfield  {title}
  {\bibinfo {title} {{Electric Field Breakdown in Single Molecule Junctions}},\
  }\href {https://doi.org/10.1021/ja512523r} {\bibfield  {journal} {\bibinfo
  {journal} {Journal of the American Chemical Society}\ }\textbf {\bibinfo
  {volume} {137}},\ \bibinfo {pages} {5028} (\bibinfo {year}
  {2015})}\BibitemShut {NoStop}%
\bibitem [{\citenamefont {Li}\ \emph {et~al.}(2016)\citenamefont {Li},
  \citenamefont {Kim}, \citenamefont {Su}, \citenamefont {Steigerwald},
  \citenamefont {Nuckolls}, \citenamefont {Darancet}, \citenamefont
  {Leighton},\ and\ \citenamefont {Venkataraman}}]{Haixing16}%
  \BibitemOpen
  \bibfield  {author} {\bibinfo {author} {\bibfnamefont {H.}~\bibnamefont
  {Li}}, \bibinfo {author} {\bibfnamefont {N.~T.}\ \bibnamefont {Kim}},
  \bibinfo {author} {\bibfnamefont {T.~A.}\ \bibnamefont {Su}}, \bibinfo
  {author} {\bibfnamefont {M.~L.}\ \bibnamefont {Steigerwald}}, \bibinfo
  {author} {\bibfnamefont {C.}~\bibnamefont {Nuckolls}}, \bibinfo {author}
  {\bibfnamefont {P.}~\bibnamefont {Darancet}}, \bibinfo {author}
  {\bibfnamefont {J.~L.}\ \bibnamefont {Leighton}},\ and\ \bibinfo {author}
  {\bibfnamefont {L.}~\bibnamefont {Venkataraman}},\ }\bibfield  {title}
  {\bibinfo {title} {{Mechanism for Si--Si Bond Rupture in Single Molecule
  Junctions}},\ }\href {https://doi.org/10.1021/jacs.6b10700} {\bibfield
  {journal} {\bibinfo  {journal} {Journal of the American Chemical Society}\
  }\textbf {\bibinfo {volume} {138}},\ \bibinfo {pages} {16159} (\bibinfo
  {year} {2016})}\BibitemShut {NoStop}%
\bibitem [{\citenamefont {Pobelov}\ \emph {et~al.}(2017)\citenamefont
  {Pobelov}, \citenamefont {Lauritzen}, \citenamefont {Yoshida}, \citenamefont
  {Jensen}, \citenamefont {M{\'{e}}sz{\'{a}}ros}, \citenamefont {Jacobsen},
  \citenamefont {Strange}, \citenamefont {Wandlowski},\ and\ \citenamefont
  {Solomon}}]{Pobelov17}%
  \BibitemOpen
  \bibfield  {author} {\bibinfo {author} {\bibfnamefont {I.~V.}\ \bibnamefont
  {Pobelov}}, \bibinfo {author} {\bibfnamefont {K.~P.}\ \bibnamefont
  {Lauritzen}}, \bibinfo {author} {\bibfnamefont {K.}~\bibnamefont {Yoshida}},
  \bibinfo {author} {\bibfnamefont {A.}~\bibnamefont {Jensen}}, \bibinfo
  {author} {\bibfnamefont {G.}~\bibnamefont {M{\'{e}}sz{\'{a}}ros}}, \bibinfo
  {author} {\bibfnamefont {K.~W.}\ \bibnamefont {Jacobsen}}, \bibinfo {author}
  {\bibfnamefont {M.}~\bibnamefont {Strange}}, \bibinfo {author} {\bibfnamefont
  {T.}~\bibnamefont {Wandlowski}},\ and\ \bibinfo {author} {\bibfnamefont
  {G.~C.}\ \bibnamefont {Solomon}},\ }\bibfield  {title} {\bibinfo {title}
  {{Dynamic breaking of a single gold bond}},\ }\href
  {http://dx.doi.org/10.1038/ncomms15931} {\bibfield  {journal} {\bibinfo
  {journal} {Nature Communications}\ }\textbf {\bibinfo {volume} {8}},\
  \bibinfo {pages} {15931} (\bibinfo {year} {2017})}\BibitemShut {NoStop}%
\bibitem [{\citenamefont {Ding}\ \emph {et~al.}(2016)\citenamefont {Ding},
  \citenamefont {Xiong},\ and\ \citenamefont {Dong}}]{Ding2016}%
  \BibitemOpen
  \bibfield  {author} {\bibinfo {author} {\bibfnamefont {G.-H.}\ \bibnamefont
  {Ding}}, \bibinfo {author} {\bibfnamefont {B.}~\bibnamefont {Xiong}},\ and\
  \bibinfo {author} {\bibfnamefont {B.}~\bibnamefont {Dong}},\ }\bibfield
  {title} {\bibinfo {title} {{Transient currents of a single molecular junction
  with a vibrational mode}},\ }\href
  {https://doi.org/10.1088/0953-8984/28/6/065301} {\bibfield  {journal}
  {\bibinfo  {journal} {Journal of Physics: Condensed Matter}\ }\textbf
  {\bibinfo {volume} {28}},\ \bibinfo {pages} {65301} (\bibinfo {year}
  {2016})}\BibitemShut {NoStop}%
\bibitem [{\citenamefont {Erpenbeck}\ \emph {et~al.}(2016)\citenamefont
  {Erpenbeck}, \citenamefont {H{\"{a}}rtle}, \citenamefont {Bockstedte},\ and\
  \citenamefont {Thoss}}]{Erpenbeck16}%
  \BibitemOpen
  \bibfield  {author} {\bibinfo {author} {\bibfnamefont {A.}~\bibnamefont
  {Erpenbeck}}, \bibinfo {author} {\bibfnamefont {R.}~\bibnamefont
  {H{\"{a}}rtle}}, \bibinfo {author} {\bibfnamefont {M.}~\bibnamefont
  {Bockstedte}},\ and\ \bibinfo {author} {\bibfnamefont {M.}~\bibnamefont
  {Thoss}},\ }\bibfield  {title} {\bibinfo {title} {{Vibrationally dependent
  electron-electron interactions in resonant electron transport through
  single-molecule junctions}},\ }\href
  {https://doi.org/10.1103/PhysRevB.93.115421} {\bibfield  {journal} {\bibinfo
  {journal} {Physical Review B}\ }\textbf {\bibinfo {volume} {93}},\ \bibinfo
  {pages} {115421} (\bibinfo {year} {2016})}\BibitemShut {NoStop}%
\bibitem [{\citenamefont {Schinabeck}\ \emph {et~al.}(2014)\citenamefont
  {Schinabeck}, \citenamefont {H{\"{a}}rtle}, \citenamefont {Weber},\ and\
  \citenamefont {Thoss}}]{thoss14}%
  \BibitemOpen
  \bibfield  {author} {\bibinfo {author} {\bibfnamefont {C.}~\bibnamefont
  {Schinabeck}}, \bibinfo {author} {\bibfnamefont {R.}~\bibnamefont
  {H{\"{a}}rtle}}, \bibinfo {author} {\bibfnamefont {H.~B.}\ \bibnamefont
  {Weber}},\ and\ \bibinfo {author} {\bibfnamefont {M.}~\bibnamefont {Thoss}},\
  }\bibfield  {title} {\bibinfo {title} {{Current noise in single-molecule
  junctions induced by electronic-vibrational coupling}},\ }\href
  {https://doi.org/10.1103/PhysRevB.90.075409} {\bibfield  {journal} {\bibinfo
  {journal} {Phys. Rev. B}\ }\textbf {\bibinfo {volume} {90}},\ \bibinfo
  {pages} {75409} (\bibinfo {year} {2014})}\BibitemShut {NoStop}%
\bibitem [{\citenamefont {Haupt}\ \emph {et~al.}(2010)\citenamefont {Haupt},
  \citenamefont {Novotn{\'{y}}},\ and\ \citenamefont {Belzig}}]{belzig10}%
  \BibitemOpen
  \bibfield  {author} {\bibinfo {author} {\bibfnamefont {F.}~\bibnamefont
  {Haupt}}, \bibinfo {author} {\bibfnamefont {T.~c.~{\v{s}}.}\ \bibnamefont
  {Novotn{\'{y}}}},\ and\ \bibinfo {author} {\bibfnamefont {W.}~\bibnamefont
  {Belzig}},\ }\bibfield  {title} {\bibinfo {title} {{Current noise in
  molecular junctions: Effects of the electron-phonon interaction}},\ }\href
  {https://doi.org/10.1103/PhysRevB.82.165441} {\bibfield  {journal} {\bibinfo
  {journal} {Phys. Rev. B}\ }\textbf {\bibinfo {volume} {82}},\ \bibinfo
  {pages} {165441} (\bibinfo {year} {2010})}\BibitemShut {NoStop}%
\bibitem [{\citenamefont {Galperin}\ \emph {et~al.}(2009)\citenamefont
  {Galperin}, \citenamefont {Saito}, \citenamefont {Balatsky},\ and\
  \citenamefont {Nitzan}}]{galperin09}%
  \BibitemOpen
  \bibfield  {author} {\bibinfo {author} {\bibfnamefont {M.}~\bibnamefont
  {Galperin}}, \bibinfo {author} {\bibfnamefont {K.}~\bibnamefont {Saito}},
  \bibinfo {author} {\bibfnamefont {A.~V.}\ \bibnamefont {Balatsky}},\ and\
  \bibinfo {author} {\bibfnamefont {A.}~\bibnamefont {Nitzan}},\ }\bibfield
  {title} {\bibinfo {title} {{Cooling mechanisms in molecular conduction
  junctions}},\ }\href {https://doi.org/10.1103/PhysRevB.80.115427} {\bibfield
  {journal} {\bibinfo  {journal} {Phys. Rev. B}\ }\textbf {\bibinfo {volume}
  {80}},\ \bibinfo {pages} {115427} (\bibinfo {year} {2009})}\BibitemShut
  {NoStop}%
\bibitem [{\citenamefont {Erpenbeck}\ and\ \citenamefont
  {Thoss}(2019)}]{Erpenbeck2019}%
  \BibitemOpen
  \bibfield  {author} {\bibinfo {author} {\bibfnamefont {A.}~\bibnamefont
  {Erpenbeck}}\ and\ \bibinfo {author} {\bibfnamefont {M.}~\bibnamefont
  {Thoss}},\ }\bibfield  {title} {\bibinfo {title} {{Hierarchical quantum
  master equation approach to vibronic reaction dynamics at metal surfaces}},\
  }\href {https://doi.org/10.1063/1.5128206} {\bibfield  {journal} {\bibinfo
  {journal} {Journal of Chemical Physics}\ }\textbf {\bibinfo {volume} {151}},\
  \bibinfo {pages} {191101} (\bibinfo {year} {2019})}\BibitemShut {NoStop}%
\bibitem [{\citenamefont {Koch}\ \emph {et~al.}(2006)\citenamefont {Koch},
  \citenamefont {Semmelhack}, \citenamefont {von Oppen},\ and\ \citenamefont
  {Nitzan}}]{koch06}%
  \BibitemOpen
  \bibfield  {author} {\bibinfo {author} {\bibfnamefont {J.}~\bibnamefont
  {Koch}}, \bibinfo {author} {\bibfnamefont {M.}~\bibnamefont {Semmelhack}},
  \bibinfo {author} {\bibfnamefont {F.}~\bibnamefont {von Oppen}},\ and\
  \bibinfo {author} {\bibfnamefont {A.}~\bibnamefont {Nitzan}},\ }\bibfield
  {title} {\bibinfo {title} {{Current-induced nonequilibrium vibrations in
  single-molecule devices}},\ }\href
  {https://doi.org/10.1103/PhysRevB.73.155306} {\bibfield  {journal} {\bibinfo
  {journal} {Phys. Rev. B}\ }\textbf {\bibinfo {volume} {73}},\ \bibinfo
  {pages} {155306} (\bibinfo {year} {2006})}\BibitemShut {NoStop}%
\bibitem [{\citenamefont {Wilner}\ \emph {et~al.}(2014)\citenamefont {Wilner},
  \citenamefont {Wang}, \citenamefont {Thoss},\ and\ \citenamefont
  {Rabani}}]{rabani14}%
  \BibitemOpen
  \bibfield  {author} {\bibinfo {author} {\bibfnamefont {E.~Y.}\ \bibnamefont
  {Wilner}}, \bibinfo {author} {\bibfnamefont {H.}~\bibnamefont {Wang}},
  \bibinfo {author} {\bibfnamefont {M.}~\bibnamefont {Thoss}},\ and\ \bibinfo
  {author} {\bibfnamefont {E.}~\bibnamefont {Rabani}},\ }\bibfield  {title}
  {\bibinfo {title} {{Nonequilibrium quantum systems with electron-phonon
  interactions: Transient dynamics and approach to steady state}},\ }\href
  {https://doi.org/10.1103/PhysRevB.89.205129} {\bibfield  {journal} {\bibinfo
  {journal} {Phys. Rev. B}\ }\textbf {\bibinfo {volume} {89}},\ \bibinfo
  {pages} {205129} (\bibinfo {year} {2014})}\BibitemShut {NoStop}%
\bibitem [{\citenamefont {Wang}\ \emph {et~al.}(2011)\citenamefont {Wang},
  \citenamefont {Pshenichnyuk}, \citenamefont {H{\"{a}}rtle},\ and\
  \citenamefont {Thoss}}]{wang11}%
  \BibitemOpen
  \bibfield  {author} {\bibinfo {author} {\bibfnamefont {H.}~\bibnamefont
  {Wang}}, \bibinfo {author} {\bibfnamefont {I.}~\bibnamefont {Pshenichnyuk}},
  \bibinfo {author} {\bibfnamefont {R.}~\bibnamefont {H{\"{a}}rtle}},\ and\
  \bibinfo {author} {\bibfnamefont {M.}~\bibnamefont {Thoss}},\ }\bibfield
  {title} {\bibinfo {title} {{Numerically exact, time-dependent treatment of
  vibrationally coupled electron transport in single-molecule junctions}},\
  }\href {https://doi.org/10.1063/1.3660206} {\bibfield  {journal} {\bibinfo
  {journal} {J. Chem. Phys.}\ }\textbf {\bibinfo {volume} {135}},\ \bibinfo
  {pages} {244506} (\bibinfo {year} {2011})}\BibitemShut {NoStop}%
\bibitem [{\citenamefont {Schinabeck}\ \emph {et~al.}(2016)\citenamefont
  {Schinabeck}, \citenamefont {Erpenbeck}, \citenamefont {H{\"{a}}rtle},\ and\
  \citenamefont {Thoss}}]{Schinabeck16}%
  \BibitemOpen
  \bibfield  {author} {\bibinfo {author} {\bibfnamefont {C.}~\bibnamefont
  {Schinabeck}}, \bibinfo {author} {\bibfnamefont {A.}~\bibnamefont
  {Erpenbeck}}, \bibinfo {author} {\bibfnamefont {R.}~\bibnamefont
  {H{\"{a}}rtle}},\ and\ \bibinfo {author} {\bibfnamefont {M.}~\bibnamefont
  {Thoss}},\ }\bibfield  {title} {\bibinfo {title} {{Hierarchical quantum
  master equation approach to electronic-vibrational coupling in nonequilibrium
  transport through nanosystems}},\ }\href
  {https://doi.org/10.1103/PhysRevB.94.201407} {\bibfield  {journal} {\bibinfo
  {journal} {Phys. Rev. B}\ }\textbf {\bibinfo {volume} {94}},\ \bibinfo
  {pages} {201407} (\bibinfo {year} {2016})}\BibitemShut {NoStop}%
\bibitem [{\citenamefont {Preston}\ \emph
  {et~al.}(2020{\natexlab{a}})\citenamefont {Preston}, \citenamefont
  {Kershaw},\ and\ \citenamefont {Kosov}}]{preston2020}%
  \BibitemOpen
  \bibfield  {author} {\bibinfo {author} {\bibfnamefont {R.~J.}\ \bibnamefont
  {Preston}}, \bibinfo {author} {\bibfnamefont {V.~F.}\ \bibnamefont
  {Kershaw}},\ and\ \bibinfo {author} {\bibfnamefont {D.~S.}\ \bibnamefont
  {Kosov}},\ }\bibfield  {title} {\bibinfo {title} {{Current-induced atomic
  motion, structural instabilities, and negative temperatures on
  molecule-electrode interfaces in electronic junctions}},\ }\href
  {https://doi.org/10.1103/PhysRevB.101.155415} {\bibfield  {journal} {\bibinfo
   {journal} {Phys. Rev. B}\ }\textbf {\bibinfo {volume} {101}},\ \bibinfo
  {pages} {155415} (\bibinfo {year} {2020}{\natexlab{a}})}\BibitemShut
  {NoStop}%
\bibitem [{\citenamefont {Preston}\ \emph
  {et~al.}(2020{\natexlab{b}})\citenamefont {Preston}, \citenamefont
  {Honeychurch},\ and\ \citenamefont {Kosov}}]{prestonAC2020}%
  \BibitemOpen
  \bibfield  {author} {\bibinfo {author} {\bibfnamefont {R.~J.}\ \bibnamefont
  {Preston}}, \bibinfo {author} {\bibfnamefont {T.~D.}\ \bibnamefont
  {Honeychurch}},\ and\ \bibinfo {author} {\bibfnamefont {D.~S.}\ \bibnamefont
  {Kosov}},\ }\bibfield  {title} {\bibinfo {title} {{Cooling molecular
  electronic junctions by AC current}},\ }\href
  {https://doi.org/10.1063/5.0019178} {\bibfield  {journal} {\bibinfo
  {journal} {Journal of Chemical Physics}\ }\textbf {\bibinfo {volume} {153}},\
  \bibinfo {pages} {121102} (\bibinfo {year} {2020}{\natexlab{b}})}\BibitemShut
  {NoStop}%
\bibitem [{\citenamefont {Kershaw}\ and\ \citenamefont
  {Kosov}(2020)}]{kershaw20}%
  \BibitemOpen
  \bibfield  {author} {\bibinfo {author} {\bibfnamefont {V.~F.}\ \bibnamefont
  {Kershaw}}\ and\ \bibinfo {author} {\bibfnamefont {D.~S.}\ \bibnamefont
  {Kosov}},\ }\bibfield  {title} {\bibinfo {title} {{Non-adiabatic effects of
  nuclear motion in quantum transport of electrons: A self-consistent
  Keldysh–Langevin study}},\ }\href {https://doi.org/10.1063/5.0023275}
  {\bibfield  {journal} {\bibinfo  {journal} {The Journal of Chemical Physics}\
  }\textbf {\bibinfo {volume} {153}},\ \bibinfo {pages} {154101} (\bibinfo
  {year} {2020})}\BibitemShut {NoStop}%
\bibitem [{\citenamefont {Micchi}\ \emph {et~al.}(2016)\citenamefont {Micchi},
  \citenamefont {Avriller},\ and\ \citenamefont {Pistolesi}}]{pistolesi16}%
  \BibitemOpen
  \bibfield  {author} {\bibinfo {author} {\bibfnamefont {G.}~\bibnamefont
  {Micchi}}, \bibinfo {author} {\bibfnamefont {R.}~\bibnamefont {Avriller}},\
  and\ \bibinfo {author} {\bibfnamefont {F.}~\bibnamefont {Pistolesi}},\
  }\bibfield  {title} {\bibinfo {title} {{Electromechanical transition in
  quantum dots}},\ }\href {https://doi.org/10.1103/PhysRevB.94.125417}
  {\bibfield  {journal} {\bibinfo  {journal} {Phys. Rev. B}\ }\textbf {\bibinfo
  {volume} {94}},\ \bibinfo {pages} {125417} (\bibinfo {year}
  {2016})}\BibitemShut {NoStop}%
\bibitem [{\citenamefont {Lu}\ \emph {et~al.}(2010)\citenamefont {Lu},
  \citenamefont {Brandbyge},\ and\ \citenamefont {Hedegard}}]{fuse}%
  \BibitemOpen
  \bibfield  {author} {\bibinfo {author} {\bibfnamefont {J.-T.}\ \bibnamefont
  {Lu}}, \bibinfo {author} {\bibfnamefont {M.}~\bibnamefont {Brandbyge}},\ and\
  \bibinfo {author} {\bibfnamefont {P.}~\bibnamefont {Hedegard}},\ }\bibfield
  {title} {\bibinfo {title} {{Blowing the Fuse: Berry's Phase and Runaway
  Vibrations in Molecular Conductors}},\ }\href
  {https://doi.org/10.1021/nl904233u} {\bibfield  {journal} {\bibinfo
  {journal} {Nano Letters}\ }\textbf {\bibinfo {volume} {10}},\ \bibinfo
  {pages} {1657} (\bibinfo {year} {2010})}\BibitemShut {NoStop}%
\bibitem [{\citenamefont {L{\"{u}}}\ \emph {et~al.}(2015)\citenamefont
  {L{\"{u}}}, \citenamefont {Christensen}, \citenamefont {Wang}, \citenamefont
  {Hedeg{\aa}rd},\ and\ \citenamefont {Brandbyge}}]{Lu15}%
  \BibitemOpen
  \bibfield  {author} {\bibinfo {author} {\bibfnamefont {J.-T.}\ \bibnamefont
  {L{\"{u}}}}, \bibinfo {author} {\bibfnamefont {R.~B.}\ \bibnamefont
  {Christensen}}, \bibinfo {author} {\bibfnamefont {J.-S.}\ \bibnamefont
  {Wang}}, \bibinfo {author} {\bibfnamefont {P.}~\bibnamefont {Hedeg{\aa}rd}},\
  and\ \bibinfo {author} {\bibfnamefont {M.}~\bibnamefont {Brandbyge}},\
  }\bibfield  {title} {\bibinfo {title} {{Current-Induced Forces and Hot Spots
  in Biased Nanojunctions}},\ }\href
  {https://doi.org/10.1103/PhysRevLett.114.096801} {\bibfield  {journal}
  {\bibinfo  {journal} {Physical Review Letters}\ }\textbf {\bibinfo {volume}
  {114}},\ \bibinfo {pages} {096801} (\bibinfo {year} {2015})}\BibitemShut
  {NoStop}%
\bibitem [{\citenamefont {Erpenbeck}\ \emph {et~al.}(2018)\citenamefont
  {Erpenbeck}, \citenamefont {Schinabeck}, \citenamefont {Peskin},\ and\
  \citenamefont {Thoss}}]{peskin18}%
  \BibitemOpen
  \bibfield  {author} {\bibinfo {author} {\bibfnamefont {A.}~\bibnamefont
  {Erpenbeck}}, \bibinfo {author} {\bibfnamefont {C.}~\bibnamefont
  {Schinabeck}}, \bibinfo {author} {\bibfnamefont {U.}~\bibnamefont {Peskin}},\
  and\ \bibinfo {author} {\bibfnamefont {M.}~\bibnamefont {Thoss}},\ }\bibfield
   {title} {\bibinfo {title} {{Current-induced bond rupture in single-molecule
  junctions}},\ }\href {https://doi.org/10.1103/PhysRevB.97.235452} {\bibfield
  {journal} {\bibinfo  {journal} {Phys. Rev. B}\ }\textbf {\bibinfo {volume}
  {97}},\ \bibinfo {pages} {235452} (\bibinfo {year} {2018})}\BibitemShut
  {NoStop}%
\bibitem [{\citenamefont {Dzhioev}\ \emph {et~al.}(2013)\citenamefont
  {Dzhioev}, \citenamefont {Kosov},\ and\ \citenamefont {von
  Oppen}}]{catalysis12}%
  \BibitemOpen
  \bibfield  {author} {\bibinfo {author} {\bibfnamefont {A.~A.}\ \bibnamefont
  {Dzhioev}}, \bibinfo {author} {\bibfnamefont {D.~S.}\ \bibnamefont {Kosov}},\
  and\ \bibinfo {author} {\bibfnamefont {F.}~\bibnamefont {von Oppen}},\
  }\bibfield  {title} {\bibinfo {title} {{Out-of-equilibrium catalysis of
  chemical reactions by electronic tunnel currents}},\ }\href
  {https://doi.org/10.1063/1.4797495} {\bibfield  {journal} {\bibinfo
  {journal} {J. Chem. Phys.}\ }\textbf {\bibinfo {volume} {138}},\ \bibinfo
  {pages} {134103} (\bibinfo {year} {2013})}\BibitemShut {NoStop}%
\bibitem [{\citenamefont {Dzhioev}\ and\ \citenamefont
  {Kosov}(2011)}]{dzhioev11}%
  \BibitemOpen
  \bibfield  {author} {\bibinfo {author} {\bibfnamefont {A.~A.}\ \bibnamefont
  {Dzhioev}}\ and\ \bibinfo {author} {\bibfnamefont {D.~S.}\ \bibnamefont
  {Kosov}},\ }\bibfield  {title} {\bibinfo {title} {{Kramers problem for
  nonequilibrium current-induced chemical reactions}},\ }\href
  {https://doi.org/10.1063/1.3626521} {\bibfield  {journal} {\bibinfo
  {journal} {J. Chem. Phys.}\ }\textbf {\bibinfo {volume} {135}},\ \bibinfo
  {pages} {74701} (\bibinfo {year} {2011})}\BibitemShut {NoStop}%
\bibitem [{\citenamefont {Preston}\ \emph {et~al.}(2021)\citenamefont
  {Preston}, \citenamefont {Gelin},\ and\ \citenamefont {Kosov}}]{preston21}%
  \BibitemOpen
  \bibfield  {author} {\bibinfo {author} {\bibfnamefont {R.~J.}\ \bibnamefont
  {Preston}}, \bibinfo {author} {\bibfnamefont {M.~F.}\ \bibnamefont {Gelin}},\
  and\ \bibinfo {author} {\bibfnamefont {D.~S.}\ \bibnamefont {Kosov}},\
  }\bibfield  {title} {\bibinfo {title} {{First-passage time theory of
  activated rate chemical processes in electronic molecular junctions}},\
  }\href {https://doi.org/10.1063/5.0045652} {\bibfield  {journal} {\bibinfo
  {journal} {The Journal of Chemical Physics}\ }\textbf {\bibinfo {volume}
  {154}},\ \bibinfo {pages} {114108} (\bibinfo {year} {2021})}\BibitemShut
  {NoStop}%
\bibitem [{\citenamefont {Dou}\ \emph {et~al.}(2017)\citenamefont {Dou},
  \citenamefont {Miao},\ and\ \citenamefont {Subotnik}}]{Dou17}%
  \BibitemOpen
  \bibfield  {author} {\bibinfo {author} {\bibfnamefont {W.}~\bibnamefont
  {Dou}}, \bibinfo {author} {\bibfnamefont {G.}~\bibnamefont {Miao}},\ and\
  \bibinfo {author} {\bibfnamefont {J.~E.}\ \bibnamefont {Subotnik}},\
  }\bibfield  {title} {\bibinfo {title} {{Born-Oppenheimer Dynamics, Electronic
  Friction, and the Inclusion of Electron-Electron Interactions}},\ }\href
  {https://doi.org/10.1103/PhysRevLett.119.046001} {\bibfield  {journal}
  {\bibinfo  {journal} {Phys. Rev. Lett.}\ }\textbf {\bibinfo {volume} {119}},\
  \bibinfo {pages} {46001} (\bibinfo {year} {2017})}\BibitemShut {NoStop}%
\bibitem [{\citenamefont {Dou}\ and\ \citenamefont {Subotnik}(2018)}]{Dou2018}%
  \BibitemOpen
  \bibfield  {author} {\bibinfo {author} {\bibfnamefont {W.}~\bibnamefont
  {Dou}}\ and\ \bibinfo {author} {\bibfnamefont {J.~E.}\ \bibnamefont
  {Subotnik}},\ }\bibfield  {title} {\bibinfo {title} {{Universality of
  electronic friction. II. Equivalence of the quantum-classical Liouville
  equation approach with von Oppen's nonequilibrium Green's function methods
  out of equilibrium}},\ }\bibfield  {journal} {\bibinfo  {journal} {Physical
  Review B}\ }\textbf {\bibinfo {volume} {97}},\ \href
  {https://doi.org/10.1103/PhysRevB.97.064303} {10.1103/PhysRevB.97.064303}
  (\bibinfo {year} {2018}),\ \Eprint {https://arxiv.org/abs/1801.06108}
  {arXiv:1801.06108} \BibitemShut {NoStop}%
\bibitem [{\citenamefont {Bode}\ \emph {et~al.}(2012)\citenamefont {Bode},
  \citenamefont {Kusminskiy}, \citenamefont {Egger},\ and\ \citenamefont {von
  Oppen}}]{Bode12}%
  \BibitemOpen
  \bibfield  {author} {\bibinfo {author} {\bibfnamefont {N.}~\bibnamefont
  {Bode}}, \bibinfo {author} {\bibfnamefont {S.~V.}\ \bibnamefont
  {Kusminskiy}}, \bibinfo {author} {\bibfnamefont {R.}~\bibnamefont {Egger}},\
  and\ \bibinfo {author} {\bibfnamefont {F.}~\bibnamefont {von Oppen}},\
  }\bibfield  {title} {\bibinfo {title} {{Current-induced forces in mesoscopic
  systems: A scattering-matrix approach}},\ }\href
  {https://doi.org/10.3762/bjnano.3.15} {\bibfield  {journal} {\bibinfo
  {journal} {Beilstein Journal of Nanotechnology}\ }\textbf {\bibinfo {volume}
  {3}},\ \bibinfo {pages} {144} (\bibinfo {year} {2012})}\BibitemShut {NoStop}%
\bibitem [{\citenamefont {Chen}\ \emph
  {et~al.}(2019{\natexlab{a}})\citenamefont {Chen}, \citenamefont {Miwa},\ and\
  \citenamefont {Galperin}}]{chen19}%
  \BibitemOpen
  \bibfield  {author} {\bibinfo {author} {\bibfnamefont {F.}~\bibnamefont
  {Chen}}, \bibinfo {author} {\bibfnamefont {K.}~\bibnamefont {Miwa}},\ and\
  \bibinfo {author} {\bibfnamefont {M.}~\bibnamefont {Galperin}},\ }\bibfield
  {title} {\bibinfo {title} {{Current-Induced Forces for Nonadiabatic Molecular
  Dynamics}},\ }\href {https://doi.org/10.1021/acs.jpca.8b09251} {\bibfield
  {journal} {\bibinfo  {journal} {The Journal of Physical Chemistry A}\
  }\textbf {\bibinfo {volume} {123}},\ \bibinfo {pages} {693} (\bibinfo {year}
  {2019}{\natexlab{a}})}\BibitemShut {NoStop}%
\bibitem [{\citenamefont {Chen}\ \emph
  {et~al.}(2019{\natexlab{b}})\citenamefont {Chen}, \citenamefont {Miwa},\ and\
  \citenamefont {Galperin}}]{galperin19}%
  \BibitemOpen
  \bibfield  {author} {\bibinfo {author} {\bibfnamefont {F.}~\bibnamefont
  {Chen}}, \bibinfo {author} {\bibfnamefont {K.}~\bibnamefont {Miwa}},\ and\
  \bibinfo {author} {\bibfnamefont {M.}~\bibnamefont {Galperin}},\ }\bibfield
  {title} {\bibinfo {title} {{Electronic friction in interacting systems}},\
  }\href {https://doi.org/10.1063/1.5095425} {\bibfield  {journal} {\bibinfo
  {journal} {The Journal of Chemical Physics}\ }\textbf {\bibinfo {volume}
  {150}},\ \bibinfo {pages} {174101} (\bibinfo {year}
  {2019}{\natexlab{b}})}\BibitemShut {NoStop}%
\bibitem [{\citenamefont {L{\"{u}}}\ \emph {et~al.}(2012)\citenamefont
  {L{\"{u}}}, \citenamefont {Brandbyge}, \citenamefont {Hedeg{\aa}rd},
  \citenamefont {Todorov},\ and\ \citenamefont {Dundas}}]{Todorov12}%
  \BibitemOpen
  \bibfield  {author} {\bibinfo {author} {\bibfnamefont {J.-T.}\ \bibnamefont
  {L{\"{u}}}}, \bibinfo {author} {\bibfnamefont {M.}~\bibnamefont {Brandbyge}},
  \bibinfo {author} {\bibfnamefont {P.}~\bibnamefont {Hedeg{\aa}rd}}, \bibinfo
  {author} {\bibfnamefont {T.~N.}\ \bibnamefont {Todorov}},\ and\ \bibinfo
  {author} {\bibfnamefont {D.}~\bibnamefont {Dundas}},\ }\bibfield  {title}
  {\bibinfo {title} {{Current-induced atomic dynamics, instabilities, and Raman
  signals: Quasiclassical Langevin equation approach}},\ }\href
  {https://doi.org/10.1103/PhysRevB.85.245444} {\bibfield  {journal} {\bibinfo
  {journal} {Physical Review B}\ }\textbf {\bibinfo {volume} {85}},\ \bibinfo
  {pages} {245444} (\bibinfo {year} {2012})}\BibitemShut {NoStop}%
\bibitem [{\citenamefont {Nocera}\ \emph {et~al.}(2011)\citenamefont {Nocera},
  \citenamefont {Perroni}, \citenamefont {{Marigliano Ramaglia}},\ and\
  \citenamefont {Cataudella}}]{nocera11}%
  \BibitemOpen
  \bibfield  {author} {\bibinfo {author} {\bibfnamefont {A.}~\bibnamefont
  {Nocera}}, \bibinfo {author} {\bibfnamefont {C.~A.}\ \bibnamefont {Perroni}},
  \bibinfo {author} {\bibfnamefont {V.}~\bibnamefont {{Marigliano Ramaglia}}},\
  and\ \bibinfo {author} {\bibfnamefont {V.}~\bibnamefont {Cataudella}},\
  }\bibfield  {title} {\bibinfo {title} {{Stochastic dynamics for a single
  vibrational mode in molecular junctions}},\ }\href
  {https://doi.org/10.1103/PhysRevB.83.115420} {\bibfield  {journal} {\bibinfo
  {journal} {Physical Review B}\ }\textbf {\bibinfo {volume} {83}},\ \bibinfo
  {pages} {115420} (\bibinfo {year} {2011})}\BibitemShut {NoStop}%
\bibitem [{\citenamefont {Dundas}\ \emph {et~al.}(2009)\citenamefont {Dundas},
  \citenamefont {McEniry},\ and\ \citenamefont {Todorov}}]{dundas09}%
  \BibitemOpen
  \bibfield  {author} {\bibinfo {author} {\bibfnamefont {D.}~\bibnamefont
  {Dundas}}, \bibinfo {author} {\bibfnamefont {E.~J.}\ \bibnamefont
  {McEniry}},\ and\ \bibinfo {author} {\bibfnamefont {T.~N.}\ \bibnamefont
  {Todorov}},\ }\bibfield  {title} {\bibinfo {title} {{Current-driven atomic
  waterwheels}},\ }\href {https://doi.org/10.1038/nnano.2008.411} {\bibfield
  {journal} {\bibinfo  {journal} {Nature Nanotechnology}\ }\textbf {\bibinfo
  {volume} {4}},\ \bibinfo {pages} {99} (\bibinfo {year} {2009})}\BibitemShut
  {NoStop}%
\bibitem [{\citenamefont {Dundas}\ \emph {et~al.}(2012)\citenamefont {Dundas},
  \citenamefont {Cunningham}, \citenamefont {Buchanan}, \citenamefont
  {Terasawa}, \citenamefont {Paxton},\ and\ \citenamefont
  {Todorov}}]{dundas12}%
  \BibitemOpen
  \bibfield  {author} {\bibinfo {author} {\bibfnamefont {D.}~\bibnamefont
  {Dundas}}, \bibinfo {author} {\bibfnamefont {B.}~\bibnamefont {Cunningham}},
  \bibinfo {author} {\bibfnamefont {C.}~\bibnamefont {Buchanan}}, \bibinfo
  {author} {\bibfnamefont {A.}~\bibnamefont {Terasawa}}, \bibinfo {author}
  {\bibfnamefont {A.~T.}\ \bibnamefont {Paxton}},\ and\ \bibinfo {author}
  {\bibfnamefont {T.~N.}\ \bibnamefont {Todorov}},\ }\bibfield  {title}
  {\bibinfo {title} {{An ignition key for atomic-scale engines}},\ }\href
  {https://doi.org/10.1088/0953-8984/24/40/402203} {\bibfield  {journal}
  {\bibinfo  {journal} {Journal of Physics: Condensed Matter}\ }\textbf
  {\bibinfo {volume} {24}},\ \bibinfo {pages} {402203} (\bibinfo {year}
  {2012})}\BibitemShut {NoStop}%
\bibitem [{\citenamefont {Cunningham}\ \emph {et~al.}(2014)\citenamefont
  {Cunningham}, \citenamefont {Todorov},\ and\ \citenamefont
  {Dundas}}]{cunningham14}%
  \BibitemOpen
  \bibfield  {author} {\bibinfo {author} {\bibfnamefont {B.}~\bibnamefont
  {Cunningham}}, \bibinfo {author} {\bibfnamefont {T.~N.}\ \bibnamefont
  {Todorov}},\ and\ \bibinfo {author} {\bibfnamefont {D.}~\bibnamefont
  {Dundas}},\ }\bibfield  {title} {\bibinfo {title} {{Nonconservative dynamics
  in long atomic wires}},\ }\href {https://doi.org/10.1103/PhysRevB.90.115430}
  {\bibfield  {journal} {\bibinfo  {journal} {Physical Review B}\ }\textbf
  {\bibinfo {volume} {90}},\ \bibinfo {pages} {115430} (\bibinfo {year}
  {2014})}\BibitemShut {NoStop}%
\bibitem [{\citenamefont {L{\"{u}}}\ \emph
  {et~al.}(2011{\natexlab{a}})\citenamefont {L{\"{u}}}, \citenamefont {Gunst},
  \citenamefont {Hedeg{\aa}rd},\ and\ \citenamefont {Brandbyge}}]{Brandbyge11}%
  \BibitemOpen
  \bibfield  {author} {\bibinfo {author} {\bibfnamefont {J.-T.}\ \bibnamefont
  {L{\"{u}}}}, \bibinfo {author} {\bibfnamefont {T.}~\bibnamefont {Gunst}},
  \bibinfo {author} {\bibfnamefont {P.}~\bibnamefont {Hedeg{\aa}rd}},\ and\
  \bibinfo {author} {\bibfnamefont {M.}~\bibnamefont {Brandbyge}},\ }\bibfield
  {title} {\bibinfo {title} {{Current-induced dynamics in carbon atomic
  contacts}},\ }\href {https://doi.org/10.3762/bjnano.2.90} {\bibfield
  {journal} {\bibinfo  {journal} {Beilstein Journal of Nanotechnology}\
  }\textbf {\bibinfo {volume} {2}},\ \bibinfo {pages} {814} (\bibinfo {year}
  {2011}{\natexlab{a}})}\BibitemShut {NoStop}%
\bibitem [{\citenamefont {Christensen}\ \emph {et~al.}(2016)\citenamefont
  {Christensen}, \citenamefont {L{\"{u}}}, \citenamefont {Hedeg{\aa}rd},\ and\
  \citenamefont {Brandbyge}}]{Christensen16}%
  \BibitemOpen
  \bibfield  {author} {\bibinfo {author} {\bibfnamefont {R.~B.}\ \bibnamefont
  {Christensen}}, \bibinfo {author} {\bibfnamefont {J.-T.}\ \bibnamefont
  {L{\"{u}}}}, \bibinfo {author} {\bibfnamefont {P.}~\bibnamefont
  {Hedeg{\aa}rd}},\ and\ \bibinfo {author} {\bibfnamefont {M.}~\bibnamefont
  {Brandbyge}},\ }\bibfield  {title} {\bibinfo {title} {{Current-induced
  runaway vibrations in dehydrogenated graphene nanoribbons}},\ }\href
  {https://doi.org/10.3762/bjnano.7.8} {\bibfield  {journal} {\bibinfo
  {journal} {Beilstein Journal of Nanotechnology}\ }\textbf {\bibinfo {volume}
  {7}},\ \bibinfo {pages} {68} (\bibinfo {year} {2016})}\BibitemShut {NoStop}%
\bibitem [{\citenamefont {Hussein}\ \emph {et~al.}(2010)\citenamefont
  {Hussein}, \citenamefont {Metelmann}, \citenamefont {Zedler},\ and\
  \citenamefont {Brandes}}]{hussein10}%
  \BibitemOpen
  \bibfield  {author} {\bibinfo {author} {\bibfnamefont {R.}~\bibnamefont
  {Hussein}}, \bibinfo {author} {\bibfnamefont {A.}~\bibnamefont {Metelmann}},
  \bibinfo {author} {\bibfnamefont {P.}~\bibnamefont {Zedler}},\ and\ \bibinfo
  {author} {\bibfnamefont {T.}~\bibnamefont {Brandes}},\ }\bibfield  {title}
  {\bibinfo {title} {{Semiclassical dynamics of nanoelectromechanical
  systems}},\ }\href {https://doi.org/10.1103/PhysRevB.82.165406} {\bibfield
  {journal} {\bibinfo  {journal} {Physical Review B}\ }\textbf {\bibinfo
  {volume} {82}},\ \bibinfo {pages} {165406} (\bibinfo {year}
  {2010})}\BibitemShut {NoStop}%
\bibitem [{\citenamefont {Hopjan}\ \emph {et~al.}(2018)\citenamefont {Hopjan},
  \citenamefont {Stefanucci}, \citenamefont {Perfetto},\ and\ \citenamefont
  {Verdozzi}}]{hopjan18}%
  \BibitemOpen
  \bibfield  {author} {\bibinfo {author} {\bibfnamefont {M.}~\bibnamefont
  {Hopjan}}, \bibinfo {author} {\bibfnamefont {G.}~\bibnamefont {Stefanucci}},
  \bibinfo {author} {\bibfnamefont {E.}~\bibnamefont {Perfetto}},\ and\
  \bibinfo {author} {\bibfnamefont {C.}~\bibnamefont {Verdozzi}},\ }\bibfield
  {title} {\bibinfo {title} {{Molecular junctions and molecular motors:
  Including Coulomb repulsion in electronic friction using nonequilibrium
  Green's functions}},\ }\href {https://doi.org/10.1103/PhysRevB.98.041405}
  {\bibfield  {journal} {\bibinfo  {journal} {Physical Review B}\ }\textbf
  {\bibinfo {volume} {98}},\ \bibinfo {pages} {041405} (\bibinfo {year}
  {2018})}\BibitemShut {NoStop}%
\bibitem [{\citenamefont {Kartsev}\ \emph {et~al.}(2014)\citenamefont
  {Kartsev}, \citenamefont {Verdozzi},\ and\ \citenamefont
  {Stefanucci}}]{kartsev14}%
  \BibitemOpen
  \bibfield  {author} {\bibinfo {author} {\bibfnamefont {A.}~\bibnamefont
  {Kartsev}}, \bibinfo {author} {\bibfnamefont {C.}~\bibnamefont {Verdozzi}},\
  and\ \bibinfo {author} {\bibfnamefont {G.}~\bibnamefont {Stefanucci}},\
  }\bibfield  {title} {\bibinfo {title} {{Nonadiabatic Van der Pol oscillations
  in molecular transport}},\ }\href
  {https://doi.org/10.1140/epjb/e2013-40905-5} {\bibfield  {journal} {\bibinfo
  {journal} {The European Physical Journal B}\ }\textbf {\bibinfo {volume}
  {87}},\ \bibinfo {pages} {14} (\bibinfo {year} {2014})}\BibitemShut {NoStop}%
\bibitem [{\citenamefont {Foti}\ and\ \citenamefont
  {V{\'{a}}zquez}(2018)}]{hector18}%
  \BibitemOpen
  \bibfield  {author} {\bibinfo {author} {\bibfnamefont {G.}~\bibnamefont
  {Foti}}\ and\ \bibinfo {author} {\bibfnamefont {H.}~\bibnamefont
  {V{\'{a}}zquez}},\ }\bibfield  {title} {\bibinfo {title} {{Origin of
  Vibrational Instabilities in Molecular Wires with Separated Electronic
  States}},\ }\href {https://doi.org/10.1021/acs.jpclett.8b00940} {\bibfield
  {journal} {\bibinfo  {journal} {The Journal of Physical Chemistry Letters}\
  }\textbf {\bibinfo {volume} {9}},\ \bibinfo {pages} {2791} (\bibinfo {year}
  {2018})}\BibitemShut {NoStop}%
\bibitem [{\citenamefont {Gunst}\ \emph {et~al.}(2013)\citenamefont {Gunst},
  \citenamefont {L{\"{u}}}, \citenamefont {Hedeg{\aa}rd},\ and\ \citenamefont
  {Brandbyge}}]{Gunst13}%
  \BibitemOpen
  \bibfield  {author} {\bibinfo {author} {\bibfnamefont {T.}~\bibnamefont
  {Gunst}}, \bibinfo {author} {\bibfnamefont {J.-T.}\ \bibnamefont {L{\"{u}}}},
  \bibinfo {author} {\bibfnamefont {P.}~\bibnamefont {Hedeg{\aa}rd}},\ and\
  \bibinfo {author} {\bibfnamefont {M.}~\bibnamefont {Brandbyge}},\ }\bibfield
  {title} {\bibinfo {title} {{Phonon excitation and instabilities in biased
  graphene nanoconstrictions}},\ }\href
  {https://doi.org/10.1103/PhysRevB.88.161401} {\bibfield  {journal} {\bibinfo
  {journal} {Physical Review B}\ }\textbf {\bibinfo {volume} {88}},\ \bibinfo
  {pages} {161401} (\bibinfo {year} {2013})}\BibitemShut {NoStop}%
\bibitem [{\citenamefont {Simine}\ and\ \citenamefont
  {Segal}(2012)}]{Simine12}%
  \BibitemOpen
  \bibfield  {author} {\bibinfo {author} {\bibfnamefont {L.}~\bibnamefont
  {Simine}}\ and\ \bibinfo {author} {\bibfnamefont {D.}~\bibnamefont {Segal}},\
  }\bibfield  {title} {\bibinfo {title} {{Vibrational cooling, heating, and
  instability in molecular conducting junctions: full counting statistics
  analysis}},\ }\href {https://doi.org/10.1039/c2cp40851a} {\bibfield
  {journal} {\bibinfo  {journal} {Physical Chemistry Chemical Physics}\
  }\textbf {\bibinfo {volume} {14}},\ \bibinfo {pages} {13820} (\bibinfo {year}
  {2012})}\BibitemShut {NoStop}%
\bibitem [{\citenamefont {H{\"{a}}rtle}\ and\ \citenamefont
  {Thoss}(2011)}]{thoss11}%
  \BibitemOpen
  \bibfield  {author} {\bibinfo {author} {\bibfnamefont {R.}~\bibnamefont
  {H{\"{a}}rtle}}\ and\ \bibinfo {author} {\bibfnamefont {M.}~\bibnamefont
  {Thoss}},\ }\bibfield  {title} {\bibinfo {title} {{Vibrational instabilities
  in resonant electron transport through single-molecule junctions}},\ }\href
  {https://doi.org/10.1103/PhysRevB.83.125419} {\bibfield  {journal} {\bibinfo
  {journal} {Phys. Rev. B}\ }\textbf {\bibinfo {volume} {83}},\ \bibinfo
  {pages} {125419} (\bibinfo {year} {2011})}\BibitemShut {NoStop}%
\bibitem [{\citenamefont {L{\"{u}}}\ \emph
  {et~al.}(2011{\natexlab{b}})\citenamefont {L{\"{u}}}, \citenamefont
  {Hedeg{\aa}rd},\ and\ \citenamefont {Brandbyge}}]{Lu11}%
  \BibitemOpen
  \bibfield  {author} {\bibinfo {author} {\bibfnamefont {J.-T.}\ \bibnamefont
  {L{\"{u}}}}, \bibinfo {author} {\bibfnamefont {P.}~\bibnamefont
  {Hedeg{\aa}rd}},\ and\ \bibinfo {author} {\bibfnamefont {M.}~\bibnamefont
  {Brandbyge}},\ }\bibfield  {title} {\bibinfo {title} {{Laserlike Vibrational
  Instability in Rectifying Molecular Conductors}},\ }\href
  {https://doi.org/10.1103/PhysRevLett.107.046801} {\bibfield  {journal}
  {\bibinfo  {journal} {Physical Review Letters}\ }\textbf {\bibinfo {volume}
  {107}},\ \bibinfo {pages} {046801} (\bibinfo {year}
  {2011}{\natexlab{b}})}\BibitemShut {NoStop}%
\bibitem [{\citenamefont {Rizzi}\ \emph {et~al.}(2016)\citenamefont {Rizzi},
  \citenamefont {Todorov}, \citenamefont {Kohanoff},\ and\ \citenamefont
  {Correa}}]{rizzi16}%
  \BibitemOpen
  \bibfield  {author} {\bibinfo {author} {\bibfnamefont {V.}~\bibnamefont
  {Rizzi}}, \bibinfo {author} {\bibfnamefont {T.~N.}\ \bibnamefont {Todorov}},
  \bibinfo {author} {\bibfnamefont {J.~J.}\ \bibnamefont {Kohanoff}},\ and\
  \bibinfo {author} {\bibfnamefont {A.~A.}\ \bibnamefont {Correa}},\ }\bibfield
   {title} {\bibinfo {title} {{Electron-phonon thermalization in a scalable
  method for real-time quantum dynamics}},\ }\href
  {https://doi.org/10.1103/PhysRevB.93.024306} {\bibfield  {journal} {\bibinfo
  {journal} {Physical Review B}\ }\textbf {\bibinfo {volume} {93}},\ \bibinfo
  {pages} {024306} (\bibinfo {year} {2016})}\BibitemShut {NoStop}%
\bibitem [{\citenamefont {Nitzan}\ and\ \citenamefont
  {Galperin}(2018)}]{galperin18}%
  \BibitemOpen
  \bibfield  {author} {\bibinfo {author} {\bibfnamefont {A.}~\bibnamefont
  {Nitzan}}\ and\ \bibinfo {author} {\bibfnamefont {M.}~\bibnamefont
  {Galperin}},\ }\bibfield  {title} {\bibinfo {title} {{Kinetic Schemes in Open
  Interacting Systems}},\ }\href {https://doi.org/10.1021/acs.jpclett.8b01886}
  {\bibfield  {journal} {\bibinfo  {journal} {The Journal of Physical Chemistry
  Letters}\ }\textbf {\bibinfo {volume} {9}},\ \bibinfo {pages} {4886}
  (\bibinfo {year} {2018})}\BibitemShut {NoStop}%
\bibitem [{\citenamefont {Hyldgaard}(2012)}]{hyldgaard12}%
  \BibitemOpen
  \bibfield  {author} {\bibinfo {author} {\bibfnamefont {P.}~\bibnamefont
  {Hyldgaard}},\ }\bibfield  {title} {\bibinfo {title} {Nonequilibrium
  thermodynamics of interacting tunneling transport: variational grand
  potential, density functional formulation and nature of steady-state
  forces},\ }\href {https://doi.org/10.1088/0953-8984/24/42/424219} {\bibfield
  {journal} {\bibinfo  {journal} {Journal of Physics: Condensed Matter}\
  }\textbf {\bibinfo {volume} {24}},\ \bibinfo {pages} {424219} (\bibinfo
  {year} {2012})}\BibitemShut {NoStop}%
\bibitem [{\citenamefont {Todorov}\ \emph {et~al.}(2010)\citenamefont
  {Todorov}, \citenamefont {Dundas},\ and\ \citenamefont
  {McEniry}}]{todorov10}%
  \BibitemOpen
  \bibfield  {author} {\bibinfo {author} {\bibfnamefont {T.~N.}\ \bibnamefont
  {Todorov}}, \bibinfo {author} {\bibfnamefont {D.}~\bibnamefont {Dundas}},\
  and\ \bibinfo {author} {\bibfnamefont {E.~J.}\ \bibnamefont {McEniry}},\
  }\bibfield  {title} {\bibinfo {title} {{Nonconservative generalized
  current-induced forces}},\ }\href
  {https://doi.org/10.1103/PhysRevB.81.075416} {\bibfield  {journal} {\bibinfo
  {journal} {Phys. Rev. B}\ }\textbf {\bibinfo {volume} {81}},\ \bibinfo
  {pages} {75416} (\bibinfo {year} {2010})}\BibitemShut {NoStop}%
\bibitem [{\citenamefont {Verdozzi}\ \emph {et~al.}(2006)\citenamefont
  {Verdozzi}, \citenamefont {Stefanucci},\ and\ \citenamefont
  {Almbladh}}]{Stefanucci06}%
  \BibitemOpen
  \bibfield  {author} {\bibinfo {author} {\bibfnamefont {C.}~\bibnamefont
  {Verdozzi}}, \bibinfo {author} {\bibfnamefont {G.}~\bibnamefont
  {Stefanucci}},\ and\ \bibinfo {author} {\bibfnamefont {C.-O.}\ \bibnamefont
  {Almbladh}},\ }\bibfield  {title} {\bibinfo {title} {{Classical Nuclear
  Motion in Quantum Transport}},\ }\href
  {https://doi.org/10.1103/PhysRevLett.97.046603} {\bibfield  {journal}
  {\bibinfo  {journal} {Phys. Rev. Lett.}\ }\textbf {\bibinfo {volume} {97}},\
  \bibinfo {pages} {46603} (\bibinfo {year} {2006})}\BibitemShut {NoStop}%
\bibitem [{\citenamefont {Brandbyge}\ \emph {et~al.}(2003)\citenamefont
  {Brandbyge}, \citenamefont {Stokbro}, \citenamefont {Taylor}, \citenamefont
  {Mozos},\ and\ \citenamefont {Ordejon}}]{Brandbyge03}%
  \BibitemOpen
  \bibfield  {author} {\bibinfo {author} {\bibfnamefont {M.}~\bibnamefont
  {Brandbyge}}, \bibinfo {author} {\bibfnamefont {K.}~\bibnamefont {Stokbro}},
  \bibinfo {author} {\bibfnamefont {J.}~\bibnamefont {Taylor}}, \bibinfo
  {author} {\bibfnamefont {J.~L.}\ \bibnamefont {Mozos}},\ and\ \bibinfo
  {author} {\bibfnamefont {P.}~\bibnamefont {Ordejon}},\ }\bibfield  {title}
  {\bibinfo {title} {{Origin of current-induced forces in an atomic gold wire:
  A first-principles study}},\ }\href@noop {} {\bibfield  {journal} {\bibinfo
  {journal} {Phys. Rev. B}\ }\textbf {\bibinfo {volume} {67}},\ \bibinfo
  {pages} {193104} (\bibinfo {year} {2003})}\BibitemShut {NoStop}%
\bibitem [{\citenamefont {Metelmann}\ and\ \citenamefont
  {Brandes}(2011)}]{Metelmann11}%
  \BibitemOpen
  \bibfield  {author} {\bibinfo {author} {\bibfnamefont {A.}~\bibnamefont
  {Metelmann}}\ and\ \bibinfo {author} {\bibfnamefont {T.}~\bibnamefont
  {Brandes}},\ }\bibfield  {title} {\bibinfo {title} {{Adiabaticity in
  semiclassical nanoelectromechanical systems}},\ }\href
  {https://doi.org/10.1103/PhysRevB.84.155455} {\bibfield  {journal} {\bibinfo
  {journal} {Physical Review B}\ }\textbf {\bibinfo {volume} {84}},\ \bibinfo
  {pages} {155455} (\bibinfo {year} {2011})}\BibitemShut {NoStop}%
\bibitem [{\citenamefont {Horsfield}\ \emph
  {et~al.}(2004{\natexlab{a}})\citenamefont {Horsfield}, \citenamefont
  {Bowler}, \citenamefont {Fisher}, \citenamefont {Todorov},\ and\
  \citenamefont {Montgomery}}]{horsfield04}%
  \BibitemOpen
  \bibfield  {author} {\bibinfo {author} {\bibfnamefont {A.~P.}\ \bibnamefont
  {Horsfield}}, \bibinfo {author} {\bibfnamefont {D.~R.}\ \bibnamefont
  {Bowler}}, \bibinfo {author} {\bibfnamefont {A.~J.}\ \bibnamefont {Fisher}},
  \bibinfo {author} {\bibfnamefont {T.~N.}\ \bibnamefont {Todorov}},\ and\
  \bibinfo {author} {\bibfnamefont {M.~J.}\ \bibnamefont {Montgomery}},\
  }\bibfield  {title} {\bibinfo {title} {{Power dissipation in nanoscale
  conductors: classical, semi-classical and quantum dynamics}},\ }\href
  {https://doi.org/10.1088/0953-8984/16/21/010} {\bibfield  {journal} {\bibinfo
   {journal} {Journal of Physics: Condensed Matter}\ }\textbf {\bibinfo
  {volume} {16}},\ \bibinfo {pages} {3609} (\bibinfo {year}
  {2004}{\natexlab{a}})}\BibitemShut {NoStop}%
\bibitem [{\citenamefont {McEniry}\ \emph {et~al.}(2007)\citenamefont
  {McEniry}, \citenamefont {Bowler}, \citenamefont {Dundas}, \citenamefont
  {Horsfield}, \citenamefont {S{\'{a}}nchez},\ and\ \citenamefont
  {Todorov}}]{todorov07}%
  \BibitemOpen
  \bibfield  {author} {\bibinfo {author} {\bibfnamefont {E.~J.}\ \bibnamefont
  {McEniry}}, \bibinfo {author} {\bibfnamefont {D.~R.}\ \bibnamefont {Bowler}},
  \bibinfo {author} {\bibfnamefont {D.}~\bibnamefont {Dundas}}, \bibinfo
  {author} {\bibfnamefont {A.~P.}\ \bibnamefont {Horsfield}}, \bibinfo {author}
  {\bibfnamefont {C.~G.}\ \bibnamefont {S{\'{a}}nchez}},\ and\ \bibinfo
  {author} {\bibfnamefont {T.~N.}\ \bibnamefont {Todorov}},\ }\bibfield
  {title} {\bibinfo {title} {{Dynamical simulation of inelastic quantum
  transport}},\ }\href@noop {} {\bibfield  {journal} {\bibinfo  {journal}
  {Journal of Physics: Condensed Matter}\ }\textbf {\bibinfo {volume} {19}},\
  \bibinfo {pages} {196201} (\bibinfo {year} {2007})}\BibitemShut {NoStop}%
\bibitem [{\citenamefont {Horsfield}\ \emph
  {et~al.}(2004{\natexlab{b}})\citenamefont {Horsfield}, \citenamefont
  {Bowler}, \citenamefont {Fisher}, \citenamefont {Todorov},\ and\
  \citenamefont {S{\'{a}}nchez}}]{horsfield2_04}%
  \BibitemOpen
  \bibfield  {author} {\bibinfo {author} {\bibfnamefont {A.~P.}\ \bibnamefont
  {Horsfield}}, \bibinfo {author} {\bibfnamefont {D.~R.}\ \bibnamefont
  {Bowler}}, \bibinfo {author} {\bibfnamefont {A.~J.}\ \bibnamefont {Fisher}},
  \bibinfo {author} {\bibfnamefont {T.~N.}\ \bibnamefont {Todorov}},\ and\
  \bibinfo {author} {\bibfnamefont {C.~G.}\ \bibnamefont {S{\'{a}}nchez}},\
  }\bibfield  {title} {\bibinfo {title} {{Beyond Ehrenfest: correlated
  non-adiabatic molecular dynamics}},\ }\href
  {https://doi.org/10.1088/0953-8984/16/46/012} {\bibfield  {journal} {\bibinfo
   {journal} {Journal of Physics: Condensed Matter}\ }\textbf {\bibinfo
  {volume} {16}},\ \bibinfo {pages} {8251} (\bibinfo {year}
  {2004}{\natexlab{b}})}\BibitemShut {NoStop}%
\bibitem [{\citenamefont {Nitzan}(2006)}]{nitzanbook}%
  \BibitemOpen
  \bibfield  {author} {\bibinfo {author} {\bibfnamefont {A.}~\bibnamefont
  {Nitzan}},\ }\href {https://doi.org/10.1093/oso/9780198529798.001.0001}
  {\emph {\bibinfo {title} {{Chemical Dynamics in Condensed Phases}}}}\
  (\bibinfo  {publisher} {Oxford University Press},\ \bibinfo {year}
  {2006})\BibitemShut {NoStop}%
\bibitem [{\citenamefont {Todorov}\ \emph {et~al.}(2014)\citenamefont
  {Todorov}, \citenamefont {Dundas}, \citenamefont {L{\"{u}}}, \citenamefont
  {Brandbyge},\ and\ \citenamefont {Hedeg{\aa}rd}}]{todorov14}%
  \BibitemOpen
  \bibfield  {author} {\bibinfo {author} {\bibfnamefont {T.~N.}\ \bibnamefont
  {Todorov}}, \bibinfo {author} {\bibfnamefont {D.}~\bibnamefont {Dundas}},
  \bibinfo {author} {\bibfnamefont {J.-T.}\ \bibnamefont {L{\"{u}}}}, \bibinfo
  {author} {\bibfnamefont {M.}~\bibnamefont {Brandbyge}},\ and\ \bibinfo
  {author} {\bibfnamefont {P.}~\bibnamefont {Hedeg{\aa}rd}},\ }\bibfield
  {title} {\bibinfo {title} {{Current-induced forces: a simple derivation}},\
  }\href {https://doi.org/10.1088/0143-0807/35/6/065004} {\bibfield  {journal}
  {\bibinfo  {journal} {European Journal of Physics}\ }\textbf {\bibinfo
  {volume} {35}},\ \bibinfo {pages} {065004} (\bibinfo {year}
  {2014})}\BibitemShut {NoStop}%
\bibitem [{\citenamefont {L{\"{u}}}\ \emph {et~al.}(2019)\citenamefont
  {L{\"{u}}}, \citenamefont {Hu}, \citenamefont {Hedeg{\aa}rd},\ and\
  \citenamefont {Brandbyge}}]{Hu19}%
  \BibitemOpen
  \bibfield  {author} {\bibinfo {author} {\bibfnamefont {J.-T.}\ \bibnamefont
  {L{\"{u}}}}, \bibinfo {author} {\bibfnamefont {B.-Z.}\ \bibnamefont {Hu}},
  \bibinfo {author} {\bibfnamefont {P.}~\bibnamefont {Hedeg{\aa}rd}},\ and\
  \bibinfo {author} {\bibfnamefont {M.}~\bibnamefont {Brandbyge}},\ }\bibfield
  {title} {\bibinfo {title} {{Semi-classical generalized Langevin equation for
  equilibrium and nonequilibrium molecular dynamics simulation}},\ }\href
  {https://doi.org/10.1016/j.progsurf.2018.07.002} {\bibfield  {journal}
  {\bibinfo  {journal} {Progress in Surface Science}\ }\textbf {\bibinfo
  {volume} {94}},\ \bibinfo {pages} {21} (\bibinfo {year} {2019})}\BibitemShut
  {NoStop}%
\bibitem [{\citenamefont {Jauho}\ \emph {et~al.}(1994)\citenamefont {Jauho},
  \citenamefont {Wingreen},\ and\ \citenamefont {Meir}}]{Jauho94}%
  \BibitemOpen
  \bibfield  {author} {\bibinfo {author} {\bibfnamefont {A.~P.}\ \bibnamefont
  {Jauho}}, \bibinfo {author} {\bibfnamefont {N.~S.}\ \bibnamefont
  {Wingreen}},\ and\ \bibinfo {author} {\bibfnamefont {Y.}~\bibnamefont
  {Meir}},\ }\bibfield  {title} {\bibinfo {title} {{Time-dependent transport in
  interacting and noninteracting resonant-tunneling systems}},\ }\href@noop {}
  {\bibfield  {journal} {\bibinfo  {journal} {Phys. Rev. B}\ }\textbf {\bibinfo
  {volume} {50}},\ \bibinfo {pages} {5528} (\bibinfo {year}
  {1994})}\BibitemShut {NoStop}%
\bibitem [{\citenamefont {McQuarrie}(2008)}]{mcquarrie}%
  \BibitemOpen
  \bibfield  {author} {\bibinfo {author} {\bibfnamefont {D.~A.}\ \bibnamefont
  {McQuarrie}},\ }\href@noop {} {\emph {\bibinfo {title} {{Quantum
  Chemistry}}}},\ \bibinfo {edition} {2nd}\ ed.\ (\bibinfo  {publisher}
  {University Science Books},\ \bibinfo {year} {2008})\BibitemShut {NoStop}%
\end{thebibliography}%
\end{document}